%% file: main.tex
\documentclass[preprint,aps,pra,showpacs,floatfix,superscriptaddress]{revtex4-1}
\usepackage[utf8]{inputenc}
\usepackage[OT1]{fontenc}

\usepackage{amsmath}
\usepackage{amssymb}
\usepackage{amsfonts}
\usepackage{verbatim,float}
\usepackage{graphicx}
\usepackage{bm}
\usepackage{bbm}
\usepackage{indentfirst}
\usepackage[colorlinks=true, allcolors=blue]{hyperref}
\usepackage{feynmp}
\usepackage{indentfirst}
\usepackage[compat=1.1.0]{tikz-feynman}
\newcommand{\veps}{\varepsilon}
\newcommand{\balpha}{\bm{\alpha}}

\newcommand{\br}{\bm{r}}

\newcommand{\bp}{\bm{p}}
\newcommand{\bD}{\bm{D}}

\newcommand{\rP}{{\rm P}}
\newcommand{\be}{\begin{eqnarray}}
\newcommand{\ee}{\end{eqnarray}}
\newcommand{\lb}{\left(}
\newcommand{\rb}{\right)}

\newcommand{\non}{\nonumber \\[2mm]}

\newcommand{\bra}[1]{\ensuremath{\langle #1|}} 
\newcommand{\ket}[1]{\ensuremath{|#1\rangle}}

\usepackage{dcolumn}
\usepackage{siunitx}
\usepackage{multirow}  
\usepackage{afterpage}

\usepackage{color}
\definecolor{BLUE}{rgb}{0.0,0.0,1.0}

\usepackage{ulem}

\begin{document}
\thispagestyle{empty}

\title{ Model-QED-operator approach to relativistic calculations of the nuclear recoil effect in many-electron
atoms and ions}

\author{I.~S.~Anisimova}
\affiliation{Department of Physics, St. Petersburg State University, Universitetskaya 7/9, 199034 St. Petersburg, Russia}

\author{A.~V.~Malyshev}
\affiliation{Department of Physics, St. Petersburg State University, Universitetskaya 7/9, 199034 St. Petersburg, Russia}

\author{D.~A.~Glazov}
\affiliation{Department of Physics, St. Petersburg State University, Universitetskaya 7/9, 199034 St. Petersburg, Russia}

\author{M.~Y.~Kaygorodov}
\affiliation{Department of Physics, St. Petersburg State University, Universitetskaya 7/9, 199034 St. Petersburg, Russia}

\author{Y.~S.~Kozhedub}
\affiliation{Department of Physics, St. Petersburg State University, Universitetskaya 7/9, 199034 St. Petersburg, Russia}

\author{G.~Plunien}
\affiliation{Institut f\"ur Theoretische Physik, Technische Universit\"at Dresden,
Mommsenstra{\ss}e 13, D-01062 Dresden, Germany}

\author{V.~M.~Shabaev}
\affiliation{Department of Physics, St. Petersburg State University, Universitetskaya 7/9, 199034 St. Petersburg, Russia}

\begin{abstract}

A model-operator approach to fully relativistic calculations of the nuclear recoil effect on energy levels in many-electron atomic systems is worked out. The one-electron part of the model operator for treating the normal mass shift beyond the Breit approximation is represented by a sum of semilocal and nonlocal potentials. The latter ones are constructed by employing the diagonal and off-diagonal matrix elements rigorously evaluated for hydrogenlike ions to first order in the electron-to-nucleus mass ratio. The specific mass shift beyond the lowest-order relativistic approximation has a form which can be directly employed in calculations. The capabilities of the method are probed by comparison of its predictions with the results of \textit{ab initio} QED calculations. The proposed operator can be easily incorporated into any relativistic calculation based on the Dirac-Coulomb-Breit Hamiltonian.

\end{abstract}

\maketitle


\section{\label{sec:0} Introduction}

An accurate description of the nuclear recoil effect is substantial for the proper analysis of a large number of spectroscopic experiments aimed to measure various atomic properties such as, e.g., binding and transition energies or bound-electron $g$ factors. Obviously, this effect is most pronounced in isotope differences of the corresponding properties, see, e.g. Refs.~\cite{Elliott:1996:1031:1996:4278:join_pr, Elliott:1998:583, Schuch:2005:183003, SoriaOrts:2006:103002, Brandau:2008:073201, Kohler:2016:10246, Maass:2019:182501, Koszorus:2021:439, Sailer:2022:479, King:2022:preprint, Han:2022:033049}. The nuclear recoil leads to the so-called mass shift. Along with the field shift caused by the finite-nuclear size, these effects constitute the dominant contribution to the isotope shifts. Joint high-precision theoretical and experimental studies of the isotope differences not only allow one to determine nuclear parameters, e.g., changes in the mean-square charge radii, but also pave the way in the search of new physics~\cite{Frugiuele:2017:015011, Berengut:2018:091801, Flambaum:2018:032510, Yerokhin:2020:012502, Rehbehn:2021:L040801, Figueroa:2022:073001, Ono:2022:021033, Debierre:2020:135527, Debierre:2021:032825, Debierre:2022:preprint_01668, Debierre:2022:preprint_04868}. 

Within the $(m/M)(\alpha Z)^4 mc^2$ approximation, where $m$ and $M$ are the masses of the electron and nucleus, respectively, $\alpha$ is the fine-structure constant, and $Z$ is the nuclear charge number, the nuclear recoil effect on binding energies can be described by the relativistic mass-shift operator~\cite{Shabaev:1985:588, Shabaev:1988:69, Palmer:1987:5987, Shabaev:1998:59}. The fully relativistic theory of the nuclear recoil effect to first order in $m/M$, to all orders in $\alpha Z$, and to zeroth order in $\alpha$ can be formulated only within the framework of quantum electrodynamics (QED)~\cite{Shabaev:1985:588, Shabaev:1988:69, Shabaev:1998:59}, see also Refs.~\cite{Pachucki:1995:1854, Yelkhovsky:recoil, Adkins:2007:042508}. Despite the smallness of the nuclear-strength parameter $\alpha Z$ for light atoms, the contribution of the higher orders may nevertheless be significant even in the case of hydrogen~\cite{Yerokhin:2015:233002, Yerokhin:2016:062514}. Therefore, an accurate treatment of the nuclear recoil effect demands a nonperturbative (in $\alpha Z$) consideration. To date, the corresponding {\it ab initio} QED calculations have been performed only for few-electron systems, see, e.g. Refs.~\cite{Artemyev:1995:1884, Shabaev:1998:4235, SoriaOrts:2006:103002, Adkins:2007:042508, Malyshev:2018:085001} and references therein. The computational difficulty of the rigorous methods rapidly increases with the number of electrons, which makes them practically infeasible at larger scales. A similar problem exists for evaluation of the radiative QED corrections associated with the electron self-energy and vacuum polarization. For this reason, approximate and efficient approaches for including both the QED and recoil effects within the methods based on the Dirac-Coulomb-Breit Hamiltonian~\cite{Grant:1970:747, Desclaux:1975:31, Bratzev:1977:173, Indelicato:1992:2426, Dzuba:1996:3948, Safronova:1999:4476, Tupitsyn:2003:022511, Kozlov:2015:199, Dzuba:2017:012503, Glazov:2017:46, Saue:2020:204104} are urgent. 

For the case of the radiative QED corrections, our group suggested the model-QED-operator approach~\cite{Shabaev:2013:012513} which recently has been extended to the region of superheavy elements~\cite{Malyshev:2022:012806}. The \texttt{QEDMOD} Fortran package to generate the operator was presented in Ref.~\cite{Shabaev:2015:175:2018:69:join_pr}. This operator was successfully applied to the approximate description of the QED effects on binding and transition energies in various many-electron systems~\cite{Tupitsyn:2016:253001, Pasteka:2017:023002, Machado:2018:032517, Si:2018:012504, Muller:2018:033416, Zaytsev:2019:052504, Yerokhin:2020:042816, Shabaev:2020:052502, Tupitsyn:2020:21, Kaygorodov:2021:012819, Skripnikov:2021:201101, Kaygorodov:2022:062805}. The main goal of the present work is to design a similar approach for the QED calculations of the nuclear recoil effect on energy levels beyond the approximation corresponding to the mass-shift operator. The application of the proposed approach in combination with the standard electron-correlation methods should make possible the approximate QED treatment of this effect in systems where rigorous calculations are rather problematic at the moment. 

In view of the significant progress achieved over the past decades in the accuracy of $g$-factor measurements in Penning traps~\cite{Haffner:2000:5308, Verdu:2004:093002, Sturm:2011:023002, Wagner:2013:033003, Sturm:2013:030501_R, Kohler:2016:10246, Glazov:2019:173001, Arapoglou:2019:253001, Sailer:2022:479}, high-precision evaluation of the nuclear recoil effect in the presence of an external magnetic field becomes essential as well. The QED theory of the nuclear recoil effect on the atomic $g$ factor valid to all orders in $\alpha Z$ was elaborated in Ref.~\cite{Shabaev:2001:052104}. The corresponding \textit{ab initio} calculations have been performed for few-electron ions in Refs.~\cite{Shabaev:2002:091801, Kohler:2016:10246, Shabaev:2017:263001, Malyshev:2017:765, Shabaev:2018:032512, Aleksandrov:2018:062521, Malyshev:2020:012513, Malyshev:2020:297}. In the case of more complicated systems, the nuclear recoil effect on the bound-electron $g$ factor can be treated nowadays only within the lowest-order relativistic approximation by means of the effective four-component approach derived from the QED formalism in Ref.~\cite{Shabaev:2017:263001}. In this context, the model-operator approach developed in the present work for the nuclear recoil effect on binding energies can be considered as a first step towards the construction of a more general operator suitable for studying the bound-electron $g$ factors as well.

The paper is organized as follows. In Sec.~\ref{sec:1}, we give a brief description of the relativistic theory of the nuclear recoil effect on energy levels in atoms and ions. Sec.~\ref{sec:2} is devoted to the construction of the model operator for QED calculations of the nuclear recoil effect. In Sec.~\ref{sec:3}, the numerical results are presented in a wide range of $Z=5-100$, and the performance of the suggested approach is demonstrated by comparing its predictions with the results of \textit{ab initio} calculations. The nonperturbative (in $\alpha Z$) expressions for the one- and two-electron matrix elements, which describe the nuclear recoil effect on the binding energies, are derived in Appendix~\ref{sec:a}. In Appendix~\ref{sec:b}, these expressions are additionally transformed to make them convenient for practical calculations. Finally, Appendix~\ref{sec:c} summarizes formulas necessary for the construction of the local effective potentials employed in the tests of the model-QED operator in Sec.~\ref{sec:3}.

Relativistic units ($\hbar=1$ and $c=1$) and the Heaviside charge unit ($e^2=4\pi\alpha$, where $e<0$ is the electron charge) are used throughout the paper.

\section{\label{sec:1} QED theory of the nuclear recoil effect}

As is well known, within the nonrelativistic approximation the nuclear recoil contribution to the binding energy of a hydrogenlike atom can be found by replacing the electron mass $m$ with the reduced mass, $m_r = {mM}/{(m + M)}$. For atoms with more than one electron, this recipe is insufficient, and the two-electron part of the nuclear recoil effect has to be taken into account~\cite{Hughes:1930:694}. The lowest-order relativistic (Breit) correction of first order in $m/M$ can be obtained by employing the mass-shift Hamiltonian~\cite{Shabaev:1985:588, Shabaev:1988:69, Palmer:1987:5987}. For $N$-electron system, this operator reads as
\begin{equation}
\label{eq:H_M}
H_{\rm MS} = \frac{1}{2M}\sum_{i,j=1}^{N} \left[ \bm{p}_i\cdot \bm{p}_j -  \frac{\alpha Z}{r_i} \left( \bm{\alpha}_i+\frac{(\bm{\alpha}_i \cdot \bm{r}_i)}{r_i^2} \bm{r}_i \right)\cdot \bm{p}_j \right] \, ,
\end{equation}
where the indices $i$ and $j$ enumerate the electrons, $\bm{p}=-i\bm{\nabla}$ is the momentum operator, $\bm{r}$ is the position vector, $r=|\bm{r}|$, and $\bm{\alpha}$ are the Dirac matrices. The first term in the square brackets in Eq. (\ref{eq:H_M}) corresponds to the nonrelativistic nuclear recoil operator, whereas the second term determines the leading relativistic correction.

Following Hughes and Eckart~\cite{Hughes:1930:694}, the nuclear recoil contribution to atomic spectra is usually divided into the normal (NMS) and specific (SMS) mass shifts. Accordingly, the Hamiltonian~(\ref{eq:H_M}) can be represented as a sum
\begin{align}
\label{eq:H_M_summ}
H_{\rm MS} = H_{\rm NMS} + H_{\rm SMS} \, ,
\end{align}
where the first operator corresponds to the terms $i=j$ in Eq.~(\ref{eq:H_M}) and the second one corresponds to $i\neq j$. For further discussion, it is useful to rewrite Eq.~(\ref{eq:H_M}) in the form
\begin{align}
\label{eq:H_M_D}
H_{\rm MS} = \frac{1}{2M}\sum_{i,j=1}^N \Big[\bm{p}_i\cdot \bm{p}_j - 2\bm{D}_i(0) \cdot \bm{p}_j \Big] \, 
\end{align}
by introducing the vector operator
\begin{align}
\label{eq:D0}
\bm{D}(0) = \frac{\alpha Z}{2r} \left( \bm{\alpha}+\frac{(\bm{\alpha} \cdot \bm{r})}{r^2} \bm{r} \right).
\end{align}
The mass-shift operator $H_{\rm MS}$ yields the nuclear recoil corrections up to the order $(m/M)(\alpha Z)^{4}mc^{2}$. This operator is widely used in relativistic calculations of atomic spectra and isotope shifts, where the nuclear recoil effect is particularly significant~\cite{Tupitsyn:2003:022511, SoriaOrts:2006:103002, Korol:2007:022103, Kozhedub:2010:042513, Gaidamauskas:2011:175003, Naze:2014:1197, Zubova:2014:062512, Naze:2015:032511, Fischer:2016:182004, Zubova:2016:052502, Filippin:2017:042502, Tupitsyn:2018:022517, Gamrath:2018:38, Zubova:2019:185001, Ekman:2019:433, Zaytsev:2019:052504, Yerokhin:2020:012502, Zubova:2020:1090, Muller:2021:L020802, Si:2021:012802, Schelfhout:2021:022806, Schelfhout:2022:022805}.

The fully relativistic theory of the nuclear recoil effect on binding energies can be formulated only within the rigorous QED approach (beyond the Breit approximation). To first order in $m/M$, to all orders in $\alpha Z$, and to zeroth order in $\alpha$ the corresponding theory was developed in Refs.~\cite{Shabaev:1985:588, Shabaev:1988:69, Shabaev:1998:59}. The formalism worked out in Ref.~\cite{Shabaev:1998:59} is the most suitable for the goals of the present study. Within this formalism, the pure nuclear recoil effect is taken into account by modifying the standard QED Hamiltonian of the electron-positron field interacting with the quantized electromagnetic field and the classical Coulomb potential of the nucleus~$V$. Namely, an extra term is added to the interaction part of the QED Hamiltonian, see Ref.~\cite{Shabaev:1998:59} for details. As a result, the pure nuclear recoil effect on energy levels can be obtained on equal footing with the non-recoil QED effects, e.g., the electron self-energy and vacuum polarization, by means of the perturbation theory in the interaction representation of the Furry picture~\cite{Furry:1951:115}. A convenient approach to construct the QED perturbation series both for single and quasi-degenerate levels is provided by the two-time Green's function (TTGF) method~\cite{TTGF}. This method is employed in Appendix~\ref{sec:a} to derive the formal expressions for the matrix elements describing the nuclear recoil effect on binding energies. Within this approach, the zeroth-order one-electron wave functions $\ket{\psi_n}$ and energies $\veps_n$ are assumed to be the solutions of the Dirac equation 
\begin{equation}
\label{eq:Dirac_eq}
h^{\rm D}\ket{\psi_n} \equiv \big[ \bm{\alpha} \cdot \bm{p} + \beta m + V \big] \ket{\psi_n}=\veps_n \ket{\psi_n} \, .
\end{equation}

\begin{figure}
\begin{center}
\includegraphics[width=0.65\columnwidth]{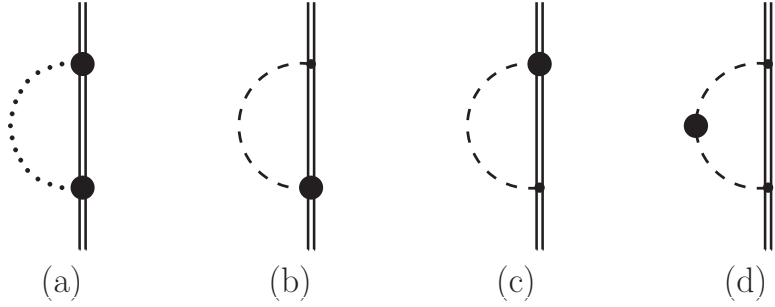}
\caption{\label{fig:recoil_1el}
One-electron nuclear recoil diagrams: the Coulomb~(a), one-transverse~(b) and (c), and two-transverse~(d) contributions. See the text and Ref.~\cite{Shabaev:1998:59} for the description of the Feynman rules.}
\end{center}
\end{figure}
\begin{figure}
\begin{center}
\includegraphics[width=0.75\columnwidth]{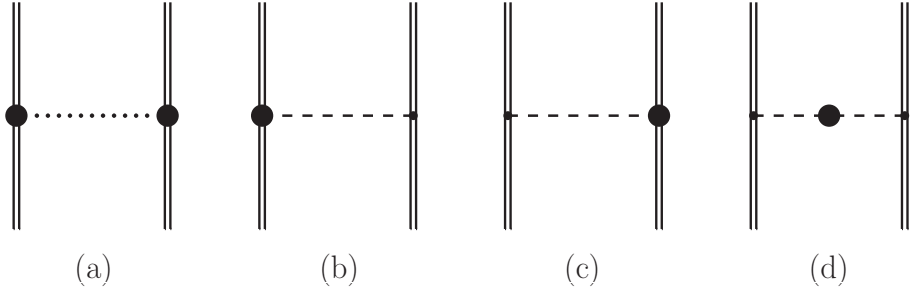}
\caption{\label{fig:recoil_2el}
Two-electron nuclear recoil diagrams: the Coulomb~(a), one-transverse~(b) and (c), and two-transverse~(d) contributions.}
\end{center}
\end{figure}

The all-order (in $\alpha Z$) expressions describing the nuclear recoil effect can be divided into the NMS and SMS parts as well. The one-electron (NMS) and two-electron (SMS) contributions are given by the Feynman diagrams shown in Figs.~\ref{fig:recoil_1el} and \ref{fig:recoil_2el}, respectively. The exhaustive description of the additional diagram-technique rules, which arise in connection with the treatment of the nuclear recoil effect, can be found in Ref.~\cite{Shabaev:1998:59}, see also Ref.~\cite{TTGF}. To explain the terminology employed throughout the paper, we briefly comment on these rules using Fig.~\ref{fig:recoil_1el} as an example. First of all, the double line and the vertex with a small dot in the figure correspond to the conventional diagram technique of the bound-state QED~\cite{TTGF}. Namely, the line denotes the electron propagator in the potential~$V$, and the vertex arises from the standard interaction of the electron-positron and electromagnetic fields. All the other diagram elements originate due to the presence of the aforementioned extra term in the QED Hamiltonian. The Coulomb gauge established itself as the most appropriate one for the nuclear-recoil-effect studies~\cite{Shabaev:1985:588, Shabaev:1988:69, Yelkhovsky:recoil, Shabaev:1998:59}, and it leads to the natural terminology. In this gauge, the photon propagator $D_{\mu\nu}(\omega,\br)$ is divided into the Coulomb,
\begin{align}
\label{eq:D_00}
D_{00}(\omega,\br) &= \frac{1}{4\pi r} \, ,
\end{align}
and transverse,
\begin{align}
\label{eq:D_lk}
D_{lk}(\omega,\br) &= 
-\frac{1}{4\pi} \Bigg[ 
\frac{\exp\left( i \sqrt{\omega^2 + i0} \, r \right)}{ r } \delta_{lk}    
+
\nabla_l \nabla_k
\frac{\exp\left( i \sqrt{\omega^2 + i0} \, r \right) - 1}{\omega^2 r}
\Bigg] \, .
\end{align}
parts, while the remaining components of the photon propagator are equal to zero, $D_{l0}=D_{0l}=0$ ($l,k=1,2,3$). All the recoil contributions in Fig.~\ref{fig:recoil_1el} can be classified with respect to the number of the propagators~(\ref{eq:D_lk}) involved. There are three possibilities~\cite{Shabaev:1998:59}: (i) the dotted line connecting two bold dots in Fig.~\ref{fig:recoil_1el}(a) depicts the so-called ``Coulomb recoil'' interaction, which does not contain the transverse part of the photon propagator at all; (ii) the dashed line with the bold dot at one of the ends in Figs.~\ref{fig:recoil_1el}(b) and \ref{fig:recoil_1el}(c) stands for the ``one-transverse-photon recoil'' interaction, which includes $D_{lk}$ once; (iii) finally, the dashed line with the bold dot in the middle in Fig.~\ref{fig:recoil_1el}(d) designates the ``two-transverse-photon recoil'' interaction, which involves the product of two photon propagators. We note that the approach initially developed in Ref.~\cite{Shabaev:1985:588} leads to the same result as the formalism of Ref.~\cite{Shabaev:1998:59} but it implies the summation of the infinite sequences of Feynman diagrams describing the electron-nucleus interaction via photon exchange. In this case, the three discussed possibilities correspond to the summation of the diagrams with zero, one, and two transverse photons and an arbitrary number of the Coulomb photons.

To all orders in $\alpha Z$, the NMS contribution for the state $|\psi_a\rangle$ can be expressed as follows~\cite{Shabaev:1985:588, Shabaev:1998:59}
\begin{equation}
\label{eq:qed_1el_P}
E_{\rm NMS}
= \langle \psi_a | {\rm P}(\veps_a) | \psi_a \rangle \, ,
\end{equation}
where we have introduced the operator ${\rm P}(E)$ by 
\begin{align}
\label{eq:P_operator}
\langle \psi_i | {\rm P}(E) | \psi_k \rangle
=
\frac{i}{2\pi} \int\limits_{-\infty}^{\infty} \! d\omega \, \sum_n 
\frac{\langle \psi_i \psi_n | R(\omega) | \psi_n \psi_k \rangle}{E - \omega - \veps_n(1-i0)} 
\end{align}
with 
\begin{align}
\label{eq:R}
R(\omega) = \frac{1}{M} \Big[ \bp_1 - \bD_1(\omega) \Big] \cdot \Big[ \bp_2 - \bD_2(\omega) \Big] \, .
\end{align}
In Eq.~(\ref{eq:P_operator}), $| \psi_i \psi_n \rangle = | \psi_i \rangle | \psi_n \rangle$ is the direct product of the one-electron wave functions. The operator $\bD(\omega)$ in Eq.~(\ref{eq:R}) is related to the transverse part of the photon propagator~(\ref{eq:D_lk}), and its $k$th Cartesian component, $D_k(\omega)$, is equal to
\begin{align}
\label{eq:D}
D_k(\omega) = -4\pi \alpha Z \alpha_l D_{lk} (\omega) \, .
\end{align}
The $\omega \rightarrow 0$ limit of the operator $\bD(\omega)$ coincides with formula~(\ref{eq:D0}). The indices 1 and 2 in Eq.~(\ref{eq:R}) designate the electron, on which the corresponding operators act. Let us rewrite $R(\omega)$ in the form:
\begin{align}
\label{eq:R_sum}
{R}(\omega) &=
{R}_{\rm c} + {R}_{\rm tr1}(\omega) + {R}_{\rm tr2}(\omega) \, ,  \\
\label{eq:R_c}
{R}_{\rm c} &= \frac{1}{M} \, \bp_1 \cdot \bp_2 \, ,  \\
\label{eq:R_tr1}
{R}_{\rm tr1}(\omega) &= -\frac{1}{M} \, \Big[ \bp_1 \cdot \bD_2(\omega) + \bD_1(\omega) \cdot \bp_2 \Big] \, ,  \\
\label{eq:R_tr2}
{R}_{\rm tr2}(\omega) &= \frac{1}{M} \, \bD_1(\omega) \cdot \bD_2(\omega) \, .
\end{align}
Substituting Eq.~(\ref{eq:R_sum}) into Eq.~(\ref{eq:qed_1el_P}), one arrives at the Coulomb, one-transverse-photon, and two-transverse-photon contributions to the NMS.

Let us turn to the discussion of the SMS contribution which corresponds to the Feynman diagrams in Fig.~\ref{fig:recoil_2el}. For simplicity, we consider the case of a one-determinant unperturbed wave function,
\begin{equation}
\label{eq:wf_2el}
\Psi_{ab}(\bm{r}_1,\bm{r}_2)=\frac{1}{\sqrt{2}} \sum_P (-1)^P \psi_{Pa}(\bm{r}_1) \psi_{Pb}(\bm{r}_2) \, ,
\end{equation}
where $P$ is the permutation operator. The generalization to the case of a many-determinant wave function is straightforward. The nonperturbative (in $\alpha Z$) expression for the SMS contribution reads as~\cite{Shabaev:1988:69, Shabaev:1998:59}
\begin{align}
\label{eq:qed_2el_R}
E_{\rm SMS}
= -\langle \psi_b \psi_a | R(\Delta) | \psi_a \psi_b \rangle \, ,
\end{align}
where $\Delta=\veps_a - \veps_b$. The formula~(\ref{eq:qed_2el_R}) gives the ``exchange'' term for the two-electron operator~$R$. The ``direct'' one is equal to zero, since the matrix elements of the operators $\bp$ and $\bD$ are zeroes for states of the same parity. Substituting (\ref{eq:R_sum}) into Eq.~(\ref{eq:qed_2el_R}), one obtains the expansion of the SMS contribution into the Coulomb, one-transverse-photon, and two-transverse-photon parts. 

Over the past three decades, numerous QED calculations of the nuclear recoil effect on binding energies were carried out~\cite{Artemyev:1995:1884, Artemyev:1995:5201, Shabaev:1998:4235, Shabaev:1999:493, SoriaOrts:2006:103002, Adkins:2007:042508, Yerokhin:2016:062514, Zubova:2016:052502, Malyshev:2018:085001, Zubova:2019:185001}. We should note that the expressions~(\ref{eq:qed_1el_P}) and (\ref{eq:qed_2el_R}) with the operator $\bD$ defined by Eq.~(\ref{eq:D}) are derived for the point-nucleus case. In Ref.~\cite{Shabaev:1998:4235}, it was argued that the dominant part of the finite-nuclear size~(FNS) correction to the nuclear recoil effect can be accounted for by employing the potential of the extended nucleus in Eq.~(\ref{eq:Dirac_eq}), and since that paper this prescription is usually used in the QED calculations of the mass shift. It was also found there that the treatment of the FNS correction to the nuclear recoil effect within the Breit approximation defined by the operator~(\ref{eq:H_M}) leads to an artificial contribution of order~$(m/M) (\alpha Z)^5 (R_{\rm nucl}/\lambdabar) mc^2$ which even exceeds the main contribution of order~$(m/M) (\alpha Z)^4 (R_{\rm nucl}/\lambdabar)^2 mc^2$ (here $\lambdabar=\hbar/(mc)$ is the Compton wavelength). This artificial contribution arises from the first (Coulomb) term in Eq.~(\ref{eq:H_M}). However, it is completely cancelled by the corresponding FNS correction to the Coulomb part of the QED nuclear recoil effect, which means that the rigorous theory for the FNS contribution beyond the main $ (m/M) (\alpha Z)^4 (R_{\rm nucl}/\lambdabar)^2 mc^2$ term can be formulated only within the framework of QED. In Ref.~\cite{Aleksandrov:2015:144004}, an additional FNS correction, which results from modifying the Breit-approximation mass-shift operator~(\ref{eq:H_M}) by inserting the form factor into the nuclear vertex~\cite{Grotch:1969:350, Borie:1982:67}, was evaluated. This operator differs from the one obtained by considering the zero-frequency limit of the photon propagator in the modified Coulomb gauge~\cite{Pachucki:2022:preprint, Veitia:2004:042501}. The main difference between the results obtained using these two operators is due to a spurious contribution of the order $(m/M) (\alpha Z)^5 (R_{\rm nucl}/\lambdabar) mc^2$ in the one-transverse-photon part, which occurs only in the calculation with the operator from Ref.~\cite{Pachucki:2022:preprint, Veitia:2004:042501}. Like to the case of the Coulomb contribution, this spurious term is cancelled by the related FNS contribution to the one-transverse-photon QED correction, provided it is also calculated with the photon propagator in the modified Coulomb gauge~\cite{Pachucki:2022:preprint, Veitia:2004:042501}. In Refs.~\cite{Yerokhin:2015:033103, Malyshev:2018:085001, Malyshev:2017:765}, the additional FNS correction from Ref.~\cite{Aleksandrov:2015:144004}, which is free from the spurious term, was used to estimate the uncertainty of the calculations based on the  prescription of Ref.~\cite{Shabaev:1998:4235}. To date, the most elaborated evaluation of the FNS correction to the nuclear recoil effect was performed within the QED approach in Ref.~\cite{Pachucki:2022:preprint}. This calculation accomplished for the $1s$ state has confirmed that the dominant part of the FNS correction to the nuclear recoil effect is indeed covered by the recipe of Ref.~\cite{Shabaev:1998:4235}, which we also follow here. The rigorous QED treatment of the total FNS correction to the nuclear recoil effect lies beyond the scope of the present work.

Restricting Eqs.~(\ref{eq:qed_1el_P}) and (\ref{eq:qed_2el_R}) by the lowest-order relativistic approximation leads to the NMS and SMS parts of the MS operator~(\ref{eq:H_M}), respectively. Until recently, all nonperturbative (in $\alpha Z$) calculations were limited by the zeroth order in $1/Z$, i.e., the electron-electron interaction corrections to the nuclear recoil effect were considered at best only within the Breit approximation. In our recent works, we have advanced the QED theory of the nuclear recoil effect. Specifically, we have considered to all orders in $\alpha Z$ the electron-electron interaction correction of first order in $1/Z$ to the one-electron~\cite{Malyshev:2019:012510} and two-electron~\cite{Malyshev:2020:052506} parts of the nuclear recoil effect on binding energies in atoms and ions. These higher-order QED contributions being also beyond the scope of the present work can be calculated additionally if needed. 

Finally, the radiative ($\sim\alpha$) as well as the second-order (in $m/M$) recoil corrections are accessible nowadays only within the $\alpha Z$-expansion approaches, see Refs.~\cite{Mohr:2012:1527, Sapirstein:1990:560, Pachucki:1995:L221} and references therein. These contributions are also not considered in this paper.

\section{\label{sec:2} Model-QED operator for the nuclear recoil effect}

\begin{figure}
\begin{center}
\includegraphics[width=0.11\columnwidth]{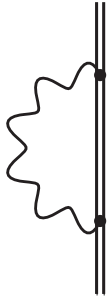}
\caption{\label{fig:se}
Self-energy diagram. 
}
\end{center}
\end{figure}
\begin{figure}
\begin{center}
\includegraphics[width=0.14\columnwidth]{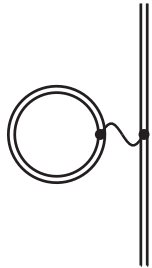}
\caption{\label{fig:vp}
Vacuum-polarization diagram. 
}
\end{center}
\end{figure}
\begin{figure}
\begin{center}
\includegraphics[width=0.17\columnwidth]{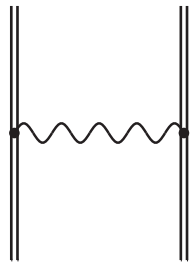}
\caption{\label{fig:1ph}
One-photon exchange diagram. 
}
\end{center}
\end{figure}

The QED calculations for many-electron systems, becoming increasingly relevant in view of the considerable progress of the experiment, are complicated and in many cases currently inaccessible. This is true for the radiative corrections as well as for the QED treatment of the nuclear recoil effect. For this reason, there is a vital need for a simple approximate approach for taking into account the QED corrections in various relativistic calculations. The conventional first-order QED corrections correspond to the self-energy (SE), vacuum-polarization (VP), and one-photon-exchange diagrams shown in Figs.~\ref{fig:se}, \ref{fig:vp}, and \ref{fig:1ph}, respectively. The approximate model-QED-operator approach to evaluate these effects has been suggested recently by our group in Refs.~\cite{Shabaev:2013:012513, Shabaev:2015:175:2018:69:join_pr, Malyshev:2022:012806}. In this section, the analogy will be traced that allows us to construct a similar approach for the nuclear recoil contributions beyond the lowest-order relativistic approximation.

The model-QED-operator approach~\cite{Shabaev:2013:012513} is worked out within the TTGF method. It is based on the fact that the QED effects to first order in $\alpha$ can be described by an effective Hamiltonian acting in the subspace which is spanned by all Slater determinants made up of the positive-energy solutions of the Dirac equation~(\ref{eq:Dirac_eq}), see Ref.~\cite{Shabaev:1993:4703} for details. This effective Hamiltonian has the form 
\begin{align}
\label{eq:H_eff}
H = \Lambda^{(+)}
\left[ 
\sum_i^N \left( h_i^{\rm D} + h_i^{\rm SE} + h_i^{\rm VP} \right)
+ 
\sum_{i<j}^N h_{ij}^{\rm 1ph}
\right]
\Lambda^{(+)}  \, ,
\end{align} 
where $\Lambda^{(+)}$ is the product of the one-electron projectors on the positive-energy eigenfunctions of the Dirac Hamiltonian~$h^{\rm D}$, and the operators~$h^{\rm SE}$, $h^{\rm VP}$, and $h^{\rm 1ph}$ arise from the diagrams in Figs.~\ref{fig:se}, \ref{fig:vp}, and \ref{fig:1ph}, respectively. According to the QED theory of the nuclear recoil effect, described briefly in Sec.~\ref{sec:1}, to first order in $m/M$ and to all orders in $\alpha Z$, the Hamiltonian $H$ has to be supplemented by the term
\begin{align}
\label{eq:H_eff_add}
\delta H = \Lambda^{(+)}
\left[ 
\sum_i^N h_i^{\rm NMS}
+ 
\sum_{i<j}^N h_{ij}^{\rm SMS}
\right]
\Lambda^{(+)}  \, ,
\end{align}
where the operators $h^{\rm NMS}$ and $h^{\rm SMS}$ are associated with the diagrams in Figs.~\ref{fig:recoil_1el} and \ref{fig:recoil_2el}, respectively. Below we discuss how the part of the operator $\delta H$ which is beyond the Breit approximation can be adapted for the practical relativistic electronic-structure calculations. We note that the entire formalism can be generalized to the case where the potential $V$ in Eq.~(\ref{eq:Dirac_eq}) along with the nuclear potential includes also some local screening potential $V_{\rm scr}$ modeling the interelectronic-interaction effects within the initial approximation, i.e., to the extended version of the Furry picture.

Let us start with the analogy between the one-photon-exchange diagram in Fig.~\ref{fig:1ph} and the two-electron nuclear recoil diagrams in Fig.~\ref{fig:recoil_2el}. For the one-photon-exchange diagram, the TTGF method leads to the following symmetric expression~\cite{Shabaev:1993:4703, TTGF}
\begin{align}
\label{eq:h_1ph}
h^{\rm 1ph} = \sum_{i_1\neq i_2, k_1\neq k_2}^{\veps_{i_1},\veps_{i_2},\veps_{k_1},\veps_{k_2}>0}
| \psi_{i_1} \psi_{i_2} \rangle 
\langle \psi_{i_1} \psi_{i_2} |
\frac{1}{2} \Big[ I(\veps_{i_1}-\veps_{k_1}) + I(\veps_{i_2}-\veps_{k_2}) \Big]
| \psi_{k_1} \psi_{k_2} \rangle 
\langle \psi_{k_1} \psi_{k_2} | \, ,
\end{align}
where the indices $i_1$, $i_2$, $k_1$, and $k_2$ enumerate the positive-energy one-electron Dirac eigenfunctions,
\begin{align}
\label{eq:I}
I(\omega) = e^2 \alpha_1^\mu \alpha_2^\nu D_{\mu\nu}(\omega,r_{12}) \, ,    
\end{align}
$\alpha^\mu\equiv\gamma^0\gamma^\mu=(1,\balpha)$ are the Dirac matrices, and the photon propagator in the Coulomb gauge is given by Eqs.~(\ref{eq:D_00}) and (\ref{eq:D_lk}). The matrix elements for the two-electron nuclear recoil diagrams are derived within the TTGF method in Appendix~\ref{sec:a}, see Eq.~(\ref{eq:H_2el_total}) for the final formula. As a result, the operator $h^{\rm SMS}$ can be represented as
\begin{align}
\label{eq:h_sms}
h^{\rm SMS} = \sum_{i_1\neq i_2, k_1\neq k_2}^{\veps_{i_1},\veps_{i_2},\veps_{k_1},\veps_{k_2}>0}
| \psi_{i_1} \psi_{i_2} \rangle 
\langle \psi_{i_1} \psi_{i_2} |
\frac{1}{2} \Big[ R(\veps_{i_1}-\veps_{k_1}) + R(\veps_{i_2}-\veps_{k_2}) \Big]
| \psi_{k_1} \psi_{k_2} \rangle 
\langle \psi_{k_1} \psi_{k_2} | \, ,
\end{align}
where the operator $R(\omega)$ is defined by Eq.~(\ref{eq:R}). Therefore, the expressions~(\ref{eq:h_1ph}) and (\ref{eq:h_sms}) differ only by the replacement of $I$ with $R$. Taking the photon propagator $D_{\mu\nu}(\omega,r_{12})$ in the Coulomb gauge at the zero-energy transfer ($\omega=0$), one obtains from Eq.~(\ref{eq:h_1ph}) the interaction part of the Dirac-Coulomb-Breit Hamiltonian. In the Coulomb gauge, the formula~(\ref{eq:h_1ph}) considered beyond the lowest-order relativistic approximation leads to the so-called frequency-dependent Breit interaction. The corresponding correction is readily taken into account within the electronic-structure calculations, see, e.g., Refs.~\cite{Chen:1997:166, Cheng:2008:052504, Yerokhin:2012:042507, Kaygorodov:2019:032505, Tupitsyn:2020:21, Kaygorodov:2022:062805}, and does not require the construction of the model operator. When omitting the two-transverse-photon contribution and taking the $\omega\rightarrow 0$ limit in $R(\omega)$, the expression~(\ref{eq:h_sms}) boils down to the SMS part of the mass-shift operator~(\ref{eq:H_M}). The remaining part of the contribution given by Eq.~(\ref{eq:h_sms}) can be treated on equal footing with the frequency-dependent Breit-interaction correction.

Now we turn to the discussion of the one-electron contributions. The VP diagram in Fig.~\ref{fig:vp} does not have a counterpart in the QED theory of the nuclear recoil effect. Therefore, we will focus on the SE diagram in Fig.~\ref{fig:se} and the one-electron nuclear recoil diagrams in Fig.~\ref{fig:recoil_1el}. First, let us introduce the SE operator~$\Sigma(E)$,
\begin{align}
\label{eq:SE_sigma}
\langle \psi_i | \Sigma(E) | \psi_k \rangle
=
\frac{i}{2\pi} \int\limits_{-\infty}^{\infty} \! d\omega \, \sum_n 
\frac{\langle \psi_i \psi_n | I(\omega) | \psi_n \psi_k \rangle}{E - \omega - \veps_n(1-i0)} \, .
\end{align} 
This formal expression suffers from ultraviolet divergences and has to be renormalized together with the mass counterterm, see, e.g., Refs.~\cite{Mohr:1974:26:1974:52:join_pr, Snyderman:1991:43, Mohr:1998:227, Yerokhin:1999:800}. Within the TTGF method, one can obtain the following symmetric expression for the operator~$h^{\rm SE}$~\cite{Shabaev:1993:4703, Shabaev:2013:012513}:
\begin{align}
\label{eq:H_se}
h^{\rm SE} =
 \sum_{i,k}^{\veps_i,\veps_k>0} 
| \psi_i \rangle \langle \psi_i | 
\frac{1}{2} \Big[ \Sigma_R(\veps_i) + \Sigma_R(\veps_k) \Big] 
| \psi_k \rangle \langle \psi_k | \, ,
\end{align} 
where $\Sigma_R(\veps)$ is the renormalized SE operator. The corresponding diagonal and off-diagonal matrix elements are tabulated, e.g., in Refs.~\cite{Shabaev:2013:012513, Malyshev:2022:012806}. The matrix elements for the one-electron nuclear recoil diagrams are considered in the framework of the TTGF method in Appendix~\ref{sec:a}, see Eq.~(\ref{eq:H_1el_total}) for the final expression. In contrast to the SE diagram, this contribution is ultraviolet finite and does not require any regularization. Nevertheless, its evaluation is a complex task compared to the calculations with the mass-shift Hamiltonian~(\ref{eq:H_M}). The operator $h^{\rm NMS}$ valid to all orders in $\alpha Z$ can be written in the form 
\begin{align}
\label{eq:H_NMS}
h^{\rm NMS} = \sum_{i,k}^{\veps_i,\veps_k>0} 
| \psi_i \rangle \langle \psi_i | 
\frac{1}{2} \Big[ \rP(\veps_i) + \rP(\veps_k) \Big] 
| \psi_k \rangle \langle \psi_k | \, ,
\end{align} 
The analogy between Eqs.~(\ref{eq:H_se}) and (\ref{eq:H_NMS}) is evident, and it is employed in the present work to develop the model-QED-operator approach for the one-electron nuclear recoil contribution. Since within the Breit approximation this contribution can be treated by means of the operator
\begin{align}
\label{eq:H_NMS_Breit}
h^{\rm NMS}_{\rm Breit} = \frac{1}{2M} \left[ \bp^2 - 2\bD(0)\cdot\bp \right] \, ,
\end{align}
we construct the model operator for the remaining (QED) part of $h^{\rm NMS}$, namely for
\begin{align}
\label{eq:H_NMS_ho}
h^{\rm NMS}_{\rm h.o.} \equiv
 \sum_{i,k}^{\veps_i,\veps_k>0} 
| \psi_i \rangle \langle \psi_i | 
\left\{
\frac{1}{2} \Big[ \rP(\veps_i) + \rP(\veps_k) \Big]
- h^{\rm NMS}_{\rm Breit}
\right\}
| \psi_k \rangle \langle \psi_k | \, .
\end{align} 

The model-operator approach should simultaneously solve two issues. First, since there is no simple-enough procedure for the \textit{ab initio} evaluation of the $\rP$-operator matrix elements for arbitrary levels (including the continuum-spectrum levels), one has to terminate the summation in Eq.~(\ref{eq:H_NMS_ho}). Second, the short interaction range should be kept. To address these problems, similarly to Ref.~\cite{Shabaev:2013:012513}, we approximate the QED recoil operator~$h^{\rm NMS}_{\rm h.o.}$ by a sum of semilocal (with respect to $r$) and nonlocal potentials
\begin{equation}
\label{eq:H_rec}
\tilde{h}^{\rm NMS}_{\rm h.o.} = V_{\rm s.l.} + V_{\rm n.l.}  \, .
\end{equation}
The operator~$\rP$, like the operator~$\Sigma_R$, conserves the relativistic angular quantum number~$\kappa=(-1)^{j+l+1/2}(j+1/2)$, where $j$ and $l$ are the total and orbital angular momenta, respectively. For this reason, the semilocal potential can be written as
\begin{equation}
\label{eq:V_loc}
V_{\rm s.l.} = \sum_{\kappa} V_{\rm s.l.}^{\kappa} P_{{\kappa}} \, ,
\end{equation}
where the projector $P_{\kappa}$ acts only on the angular variables, and its kernel is 
\begin{equation}
P_{\kappa}(\bm{n}, \bm{n}') = \left( \begin{array}{cc} \sum\limits_{m} \Omega_{\kappa m}(\bm{n}) \Omega_{\kappa m}^{\dagger}(\bm{n}') & 0 \\ 0 &  \sum\limits_{m} \Omega_{-\kappa m}(\bm{n}) \Omega_{-\kappa m}^{\dagger}(\bm{n}') \end{array} \right) 
\end{equation}
with $\Omega_{\kappa m}(\bm{n})$ being the spherical spinor and $\bm{n}=\br/r$. In Eq.~(\ref{eq:V_loc}), we take
\begin{align}
\label{eq:V_loc_exp}
V_{\rm s.l.}^{\kappa}(r) = A_{\kappa} \exp(-r/\lambdabar) \, ,
\end{align}
where the parameters $A_{\kappa}$ are chosen to reproduce the \textit{ab initio} values of the diagonal matrix elements of the operator~$h^{\rm NMS}_{\rm h.o.}$ for the state with the lowest principal quantum number $n$ for a given $\kappa$. The nonlocal potential can be written in the form
\begin{align}
\label{eq:V_nl}
V_{\rm n.l.} = \sum_{j,l = 1}^{n} \ket{\phi_j} {B_{jl}} \bra{\phi_l}  \, ,
\end{align}
where the functions~$\{\phi_i\}_{i=1}^n$, as in Ref.~\cite{Shabaev:2013:012513}, are chosen to be
\begin{align}
\label{eq:proj}
\phi_i(\br) = \frac{1}{2} \left[ \, I - (-1)^{s_i} \beta \, \right] \rho_{l_i}(r) \psi_i(\br) \, .
\end{align}
Here the index $s_i=n_i-l_i$ enumerates the positive-energy states for the given angular symmetry, $I$ and $\beta$ are the identity and the standard Dirac matrices, respectively, and the factors $\rho_{l_i}(r) = \exp \left[ -2\alpha Z(r/\lambdabar)/(1+l_i) \right]$ provide the stronger localization of the functions $\{\phi_i\}_{i=1}^n$ as compared to the eigenfunctions~ $\{\psi_i\}_{i=1}^n$ of the Dirac equation~(\ref{eq:Dirac_eq}). Finally, the coefficients $B_{jl}$ in Eq.~(\ref{eq:V_nl}) are determined from the condition that the matrix elements of the model-QED operator~$\tilde{h}^{\rm NMS}_{\rm h.o.}$ evaluated in the space spanned by the functions $\{\psi_i\}_{i=1}^n$ coincide with the exact ones. This leads to the following equations
\begin{align}
\label{eq:condition}
\sum_{j,l}^n 
\langle \psi_i | \phi_j \rangle B_{jl} \langle \phi_l | \psi_k \rangle
=
\langle \psi_i |
\left\{
\frac{1}{2} \Big[ \rP(\veps_i) + \rP(\veps_k) \Big]
- h^{\rm NMS}_{\rm Breit} - V_{\rm s.l.}
\right\}
| \psi_k \rangle
\end{align}
for $i,k=1\ldots n$, see Ref.~\cite{Shabaev:2013:012513} for details. We note that the model-QED operator for the nuclear recoil effect is worked out in such a way that the \texttt{QEDMOD} Fortran package~\cite{Shabaev:2015:175:2018:69:join_pr} can be readily adapted to incorporate it. We propose to refer to the resulting package as the \texttt{RECMOD} one.

Therefore, to determine the model-QED operator we need to calculate the diagonal and off-diagonal matrix elements of the QED recoil operator~$h^{\rm NMS}_{\rm h.o.}$ with the Dirac-Coulomb wave functions. So far, only calculations of the diagonal matrix elements have been presented in the literature. In this work, the expressions derived within the TTGF method are employed for the \textit{ab initio} QED evaluation of the one-electron nuclear recoil contributions. Concluding the description of the model-QED operator, we note that in practical calculations we construct the model operator employing the functions~(\ref{eq:proj}) which correspond to the $ns$ states with the principal quantum number $n \leqslant 3$ and the $np_{1/2}$, $np_{3/2}$, $nd_{3/2}$, and $nd_{5/2}$ states with~$n \leqslant 4$.

\section{\label{sec:3} Test of the model-QED operator for the nuclear recoil effect}

 \input{ns_nondiag.tex}
 \input{np1_nondiag.tex}
 \input{np3_nondiag.tex}
 \input{nd3_nondiag.tex}
 \input{nd5_nondiag.tex}

In the present work, to obtain the diagonal and off-diagonal matrix elements of the higher-order operator~$h^{\rm NMS}_{\rm h.o.}$, we first evaluate the corresponding values for the operator~$h^{\rm NMS}$ and then numerically subtract the Breit contribution given by the operator~$h^{\rm NMS}_{\rm Breit}$. For the Coulomb part of the nuclear recoil effect, this subtraction is readily accomplished analytically by omitting the summation over the positive-energy states and doubling the contribution of the negative-energy continuum in Eq.~(\ref{eq:omint_c}). This can be useful to avoid large cancellations for low values of $Z$. The calculations are performed for the $ns$, $np_{1/2}$, $np_{3/2}$, $nd_{3/2}$, and $nd_{5/2}$ states with the principal quantum number $n \leqslant 5$ in the wide range of $Z = 5 - 100$ using the Dirac-Coulomb wave functions. The evaluation is carried out for the potential of the extended nucleus in Eq.~(\ref{eq:Dirac_eq}). The nuclear-charge distribution is described by the homogeneously-charged-sphere model for $Z\leqslant 14$, and by the Fermi model otherwise. The nuclear-charge radii are taken from Refs.~\cite{Angeli:2013:69, Yerokhin:2015:033103}. The one-electron basis set to represent the electron Green's function is constructed from B splines~\cite{Johnson:1988:307} within the dual-kinetic-balance approach~\cite{splines:DKB}. The $\omega$ integration is performed in accordance with the equations presented in Appendix~\ref{sec:b}. 

The results of the \textit{ab initio} calculations are conveniently expressed in terms of the dimensionless function $F_{n_in_k}(\alpha Z)$ defined by
\begin{align}
\label{eq:F_ik}
\langle \psi_i | h^{\rm NMS}_{\rm h.o.} | \psi_k \rangle = 
\frac{m}{M} \frac{(\alpha Z)^{5}}{(n_in_k)^{3/2}} F_{n_in_k}(\alpha Z) \,  mc^2 \, .
\end{align}
Our data for the $ns$, $np_{1/2}$, $np_{3/2}$, $nd_{3/2}$, and $nd_{5/2}$ states are presented in Tables~\ref{tab:ns_nondiag}, \ref{tab:np1_nondiag}, \ref{tab:np3_nondiag}, \ref{tab:nd3_nondiag}, and \ref{tab:nd5_nondiag}, respectively. The individual Coulomb, one- and two-transverse-photon corrections as well as the total QED recoil contributions are shown. The uncertainties due to the approximate treatment of the nuclear-size correction are omitted, and with this in mind the values are accurate to all digits quoted. The function~$F_{n_in_k}(\alpha Z)$ for values of $Z$ not listed in Tables~\ref{tab:ns_nondiag}-\ref{tab:nd5_nondiag} can be obtained using a polynomial fitting,
\begin{align}
\label{eq:interpol}
F_{n_in_k}(\alpha Z) = 
\sum_{n=1}^N F_{n_in_k}(\alpha Z_n) 
\prod_{m\neq n} \frac{Z-Z_m}{Z_n-Z_m} \, . 
\end{align} 

\input{recmod_test.tex}


The model-QED operator for the nuclear recoil effect is constructed using the matrix elements for the $ns$ states with $n \leqslant 3$ and the $np$ and $nd$ states with $n \leqslant 4$. By definition, the operator exactly reproduces the QED recoil contributions for these states. For this reason, the ``predictive power'' of the operator can be tested by applying it to evaluation of the corresponding corrections for the states with higher values of the principal quantum number. In Table~\ref{tab:recmod_predict}, the nuclear recoil contributions beyond the Breit approximation are given for the $4s$, $5s$, $5p_{1/2}$, $5p_{3/2}$, $5d_{3/2}$, and $5d_{5/2}$ states in hydrogenlike ions. The columns labeled $\langle\psi_v| V_{\rm s.l.} |\psi_v \rangle$ and $\langle\psi_v| \tilde{h}^{\rm NMS}_{\rm h.o.} |\psi_v \rangle$ contain the predictions obtained using the semilocal part~(\ref{eq:V_loc}) of the model-QED operator and the total model-QED operator~(\ref{eq:H_rec}), respectively. These values are compared with the \textit{ab initio} results taken from Tables~\ref{tab:ns_nondiag}-\ref{tab:nd5_nondiag} and shown in the last column. As one can see, there is generally good agreement between the data, especially for the $s$ states. We stress the importance of the nonlocal part of the model-QED operator. For the $np$ and $nd$ states the functions $F_{n_in_k}$ change the sign for the middle values of $Z$ and, accordingly, have small absolute values there. As a result, the relative accuracy of the model-QED-operator predictions slightly decreases in these regions. 

\input{recmod_alkali.tex}

To date, the QED contribution to the NMS in many-electron systems was usually evaluated within the independent-electron approximation by performing the calculations in the extended Furry picture, see, e.g., Refs.~\cite{SoriaOrts:2006:103002, Zubova:2016:052502, King:2022:preprint}. To demonstrate the performance of the developed model-QED-operator approach, we apply it to evaluation of the nuclear recoil effect on valence-electron energies in neutral alkali metals. First, we perform \textit{ab initio} calculations of the one-electron contribution for the $ns$ states by using a local effective potential as $V$ in Eq.~(\ref{eq:Dirac_eq}). We use the core-Hartree (CH) and $x_\alpha$ potentials, which are discussed in Appendix~\ref{sec:c}. The results of \textit{ab initio} calculations of the QED contribution to the NMS are presented in Table~\ref{tab:recmod_alkali} in rows labeled ``Exact''. The model-QED-operator values are obtained by averaging the operator~(\ref{eq:H_rec}) with the valence-electron wave functions determined from the Dirac equation~(\ref{eq:Dirac_eq}), in which the potential $V$ is given by Eq.~(\ref{eq:V}). Along with the total model-operator predictions, $\langle\psi_v| \tilde{h}^{\rm NMS}_{\rm h.o.} |\psi_v \rangle$, the results~$\langle\psi_v| V_{\rm s.l.} |\psi_v \rangle$ for the semilocal part~(\ref{eq:V_loc}) only are shown as well. As it is seen from Table~\ref{tab:recmod_alkali}, for all the alkali metals there is good agreement between the approximate and exact values. Therefore, the model-QED operator based on the calculations for hydrogenlike ions works also reasonably well in the nonhydrogenic cases. 

To conclude the discussion of alkali metals, let us note that, as in Ref.~\cite{Sapirstein:2002:042501}, the strong dependence of the final results on the choice of the potential for the initial approximation takes place. The scatter of the results is related obviously with the approximate treatment of the interelectronic-interaction effects. We believe that the developed model-QED operator for the nuclear recoil effect merged with the standard methods to treat the electron correlations will make it possible to perform much more accurate evaluation. However, such calculations are out of the scope of the present work. We reserve systematic calculations with the model-QED operator for the nuclear recoil effect for future research.

\section{Conclusion}

In the present paper, we have worked out the model-QED-operator approach to treat the nuclear recoil effect on binding energies in many-electron atomic systems beyond the Breit approximation. The approach is similar to the one proposed earlier for approximate calculations of the radiative corrections to energy levels~\cite{Shabaev:2013:012513}. The developed operator can be readily included into any relativistic calculations based on the Dirac-Coulomb-Breit Hamiltonian. The performance of the approach was demonstrated by comparing the model-QED-operator predictions with the results of the rigorous QED calculations.

\section{Acknowledgements}

The work was supported by the Foundation for the Advancement of Theoretical
Physics and Mathematics ``BASIS'' and by RFBR and ROSATOM according to the research project No. 20-21-00098. The work of I.S.A. was supported by the German-Russian Interdisciplinary Science Center (G-RISC) funded by the German Federal Foreign Office via the German Academic Exchange Service (DAAD).


\appendix

\section{\label{sec:a} One- and two-electron contributions to the nuclear recoil effect on binding energies of quasi-degenerate levels}

In the present Appendix, we derive the fully relativistic expressions for the contributions of the nuclear recoil effect on binding energies within the two-time Green's function (TTGF) method~\cite{TTGF}. Let us denote the unperturbed wave functions of the two states under consideration as $|u_i\rangle$ and $|u_k\rangle$. During the derivation, we assume that the unperturbed energies $E_i^{(0)}$ and $E_k^{(0)}$, which corresponds to these states, differ and, therefore, $|u_i\rangle \neq |u_k\rangle$. However, all the obtained expressions are valid for the coinciding energies and, of course, boil down to the expressions~(\ref{eq:qed_1el_P}) and (\ref{eq:qed_2el_R}) in the case of diagonal matrix elements. In order to derive the formulas, we introduce a model subspace $\Omega$, which is spanned by the states $|u_i\rangle$ and $|u_k\rangle$, and construct for these states the QED perturbation theory as for quasi-degenerate levels. The projector on $\Omega$ reads as
\begin{align}
\label{eq:quasi_P0}
P^{(0)} = | u_i \rangle \langle u_i | + | u_k \rangle \langle u_k | \, .
\end{align}
The derivation procedure can be readily generalized for an arbitrary number of quasi-degenerate levels.

First, let us recall the basic ideas of the TTGF methods in the application to quasi-degenerate levels. The detailed description of the method can be found, e.g., in Refs.~\cite{TTGF, Shabaev:1993:4703, Shabaev:1994:4521}. The $N$-electron TTGF is defined as
\begin{align}
\label{eq:TTGF:G}
G( t',t ; \br'_1, \ldots , \br'_N ; \br_1, \ldots , \br_N ) 
= 
\langle 0 | T \psi(x'_1) \cdots \psi(x'_N) \bar{\psi}(x_N) \cdots \bar{\psi}(x_1) | 0 \rangle 
\left
|_{\substack{ t_1^{\prime 0}=\ldots=t_N^{\prime 0}\equiv t^\prime \\
              t_1^{0}=\ldots=t_N^{0}\equiv t }}
\right.
\, ,
\end{align}
where $\psi$ is the electron-positron field operator in the Heisenberg representation, $\bar{\psi} = \psi^\dagger\gamma^0$, $x=(t^0,\br)$, and $T$~is the time-ordering operator. Turning to the mixed representation, one obtains
\begin{align}
\label{eq:G_TTGF}
\mathcal{G} (E; \br'_1, \ldots , \br'_N ; &\, \br_1, \ldots , \br_N ) \delta (E-E')   \nonumber \\
&=
\frac{1}{2\pi i} \frac{1}{N!}
\int_{-\infty}^\infty \! dt dt^\prime \,\, 
e^{iE^\prime t^\prime -iEt} \,
G( t',t ; \br'_1, \ldots , \br'_N ; \br_1, \ldots , \br_N ) \, .
\end{align}
Employing $P^{(0)}$, we can introduce the projection of the Green's function~(\ref{eq:G_TTGF}) on the subspace~$\Omega$,
\begin{align}
\label{eq:P0_G_P0}
g(E) = P^{(0)} \mathcal{G} (E) \gamma_1^0 \ldots \gamma_N^0 P^{(0)} \, ,
\end{align}
and then determine the $\hat{K}$ and $\hat{P}$ operators:
\begin{align}
\label{eq:TTGF:K}
\hat{K} &\equiv \frac{1}{2\pi i} \oint_\Gamma \! dE \,\, E g(E) \, , \\
\label{eq:TTGF:P}
\hat{P} &\equiv \frac{1}{2\pi i} \oint_\Gamma \! dE \,\,   g(E) \, .
\end{align}
The anticlockwise oriented contour~$\Gamma$ in the complex $E$ plane surrounds the poles corresponding to the quasi-degenerate levels and keeps outside all other singularities of $g(E)$. The investigated system is fully described by the effective operator~$\hat{H}$ defined as
\begin{align}
\label{eq:TTGF:H}
\hat{H} = \hat{P}^{-1/2} \, \hat{K} \, \hat{P}^{-1/2} \, . 
\end{align}
The perturbation theory for the Green's function~(\ref{eq:TTGF:G}) leads to the perturbation series for the operator~$\hat{H}$. 

To first order in $m/M$, the nuclear recoil effect is described by the one- and two-electron Feynman diagrams depicted in Figs.~\ref{fig:recoil_1el} and \ref{fig:recoil_2el}. The Feynman rules for these diagrams are formulated, e.g., in Ref.~\cite{Shabaev:1998:59}, see also Ref.~\cite{TTGF}. The first-order contribution to the operator~$\hat{H}$ can be expressed as
\begin{align}
\label{eq:H_1}
\hat{H}^{(1)} = \hat{K}^{(1)} - \frac{1}{2} \hat{P}^{(1)} \hat{K}^{(0)} - \frac{1}{2} \hat{K}^{(0)} \hat{P}^{(1)} \, ,
\end{align}
where the superscripts denote the orders in the expansion parameter. The contributions of the nuclear recoil effect are determined by the matrix elements of the operator~(\ref{eq:H_1}) in the basis of the unperturbed functions $|u_i\rangle$ and $|u_k\rangle$:
\begin{align}
\label{eq:H_1_ik}
H^{(1)}_{ik} \equiv \langle u_i | \hat{H}^{(1)} | u_k \rangle  \, .
\end{align}
To zeroth order, the matrix of the operator $\hat{K}$ is diagonal, $K^{(0)}_{ik}=E^{(0)}_i\delta_{ik}$. Therefore, one can obtain
\begin{align}
\label{eq:H_1_ik_2}
H^{(1)}_{ik} = K^{(1)}_{ik} - \frac{E^{(0)}_i+E^{(0)}_k}{2} P^{(1)}_{ik}   \, .
\end{align}

Let us now directly turn to the derivation of the desired nonperturbative (in $\alpha Z$) formulas for the nuclear recoil effect on binding energies. We start from the one-electron (NMS) contribution corresponding to the diagrams in Fig.~\ref{fig:recoil_1el}. In this case, $N=1$ and the unperturbed wave functions~$|u_i\rangle$, $|u_k\rangle$ and energies~$E^{(0)}_i$, $E^{(0)}_k$ are given by the Dirac eigenfunctions and eigenvalues, respectively, see Eq.~(\ref{eq:Dirac_eq}). The present derivation is similar to the one for the self-energy diagram shown in Fig.~\ref{fig:se}. Employing the Feynman rules~\cite{Shabaev:1998:59}, one can obtain the following expression for the matrix element of the Green's function~$\Delta g_{{\rm NMS}}^{(1)}(E)$ projected on the subspace $\Omega$~\cite{TTGF}: 
\begin{align}
\label{eq:g_1el}
\Delta g_{{\rm NMS}, ik}^{(1)}(E) = \frac{\langle \psi_i | \rP(E) | \psi_k \rangle}{(E-\veps_i)(E-\veps_k)} \, ,
\end{align}
where the operator ${\rm P}$ is defined in Eq.~(\ref{eq:P_operator}). The corresponding contributions to the $\hat{K}$ and $\hat{P}$ operators read as:
\begin{align}
\label{eq:K_1_1el}
K_{{\rm NMS}, ik}^{(1)} &= \frac{1}{2\pi i} \oint_\Gamma \! dE \, E \, \Delta g_{{\rm NMS}, ik}^{(1)}(E)  
= \frac{\veps_i}{\veps_i-\veps_k} \langle \psi_i | \rP(\veps_i) | \psi_k \rangle
 + \frac{\veps_k}{\veps_k-\veps_i} \langle \psi_i | \rP(\veps_k) | \psi_k \rangle  \, ,   \\
\label{eq:P_1_1el} 
P_{{\rm NMS}, ik}^{(1)} &= \frac{1}{2\pi i} \oint_\Gamma \! dE \, \Delta g_{{\rm NMS}, ik}^{(1)}(E)   
= \frac{1}{\veps_i-\veps_k} \langle \psi_i | \rP(\veps_i) | \psi_k \rangle
 + \frac{1}{\veps_k-\veps_i} \langle \psi_i | \rP(\veps_k) | \psi_k \rangle  \, .   
\end{align}
Substituting Eqs.~(\ref{eq:K_1_1el}) and (\ref{eq:P_1_1el}) into the formula~(\ref{eq:H_1_ik_2}), one finally obtains the NMS contribution 
\begin{align}
\label{eq:H_1el_total}
H_{{\rm NMS}, ik}
= \frac{1}{2} 
\Big[ \langle \psi_i | \rP(\veps_i) | \psi_k \rangle 
    + \langle \psi_i | \rP(\veps_k) | \psi_k \rangle \Big]  \, .
\end{align}

The derivation of the nonperturbative (in $\alpha Z$) expression for the two-electron (SMS) contribution in Fig.~\ref{fig:recoil_2el} is also straightforward but more tedious. In this case, $N=2$ and we, for simplicity, assume that the unperturbed wave functions $|u_i\rangle$ and $|u_k\rangle$ are given by the one-determinant wave functions $\Psi_{i_1i_2}$ and $\Psi_{k_1k_2}$, respectively, see Eq.~(\ref{eq:wf_2el}). The corresponding unperturbed energies are equal to $E^{(0)}_i=\veps_{i_1}+\veps_{i_2}$ and $E^{(0)}_k=\veps_{k_1}+\veps_{k_2}$. The derivation repeats the one for the one-photon-exchange diagram in Fig.~\ref{fig:1ph}. The Green's function~$\Delta g_{{\rm SMS}}^{(1)}(E)$ projected on the subspace $\Omega$ can be written as~\cite{Shabaev:1998:59, TTGF}
\begin{align}
\label{eq:g_2el}
&\Delta g^{(1)}_{{\rm SMS},ik}(E) = \lb \frac{i}{2\pi} \rb^2 \int \! dp_1 dp_1' \, \sum_P (-1)^P \,
    \langle \psi_{Pi_1} \psi_{Pi_2} | R(p_1-p_1') | \psi_{k_1} \psi_{k_2} \rangle   \non
  & \qquad \times    
    \frac{1}{(p_1'-\veps_{Pi_1}+i0)(E-p_1'-\veps_{Pi_2}+i0)}  
    \frac{1}{(p_1 -\veps_{k_1} +i0)(E-p_1 -\veps_{k_2} +i0)}   \, ,
\end{align}
where the operator $R$ is defined by Eq.~(\ref{eq:R}). Using the identity
\begin{align}
\frac{1}{(p_1'-\veps_{Pi_1}+i0)(E-p_1'-\veps_{Pi_2}+i0)} 
= 
\frac{1}{E-E_i^{(0)}} \lb \frac{1}{p_1'-\veps_{Pi_1}+i0} + \frac{1}{E-p_1'-\veps_{Pi_2}+i0} \rb 
\end{align}
and a similar one for the second pair of denominators in Eq.~(\ref{eq:g_2el}), one can explicitly separate the poles of the Green's function located inside the contour $\Gamma$. Then, the contribution to the $\hat{K}$ operator is
\begin{align}
K_{{\rm SMS}, ik}^{(1)} &=  \oint_\Gamma \! dE \, E \, \Delta g^{(1)}_{{\rm SMS},ik}(E) \non
&= 
  \lb \frac{i}{2\pi} \rb^2 \int \! dp_1 dp_1' \, \sum_P (-1)^P \,
     \langle \psi_{Pi_1} \psi_{Pi_2} | R(p_1-p_1') | \psi_{k_1} \psi_{k_2} \rangle  \non
& \times \,    
    \Bigg\{\,
      \frac{E_i^{(0)}}{E_i^{(0)}-E_k^{(0)}}
      \lb \frac{2\pi}{i} \delta( p_1' - \veps_{Pi_1} ) \rb
      \lb \frac{1}{p_1 -\veps_{k_1} +i0} + \frac{1}{E_i^{(0)}-p_1 -\veps_{k_2} +i0} \rb    \non
& \quad  + \,      
      \frac{E_k^{(0)}}{E_k^{(0)}-E_i^{(0)}}
      \lb \frac{1}{p_1' -\veps_{Pi_1} +i0} + \frac{1}{E_k^{(0)}-p_1' -\veps_{Pi_2} +i0} \rb
      \lb \frac{2\pi}{i} \delta( p_1 - \veps_{k_1} ) \rb
    \,\Bigg\} \, ,
\end{align}
where the indentity
\begin{align}
\frac{1}{p-\veps+i0} + \frac{1}{-p+\veps+i0} = \frac{2\pi}{i}\delta(p-\veps)
\end{align}
was employed. Defining the coefficients $A$ and $B$ so that $K_{{\rm SMS}, ik}^{(1)} \equiv A E_i^{(0)} + B E_k^{(0)} $, the contribution to the operator $\hat{P}$ can be expressed as $P_{{\rm SMS}, ik}^{(1)} = A + B$. Therefore, the formula~(\ref{eq:H_1_ik_2}) results in
\begin{align}
\label{eq:H_2el}
H_{{\rm SMS}, ik}
&= \frac{1}{2} \, \frac{i}{2\pi} \, \int \! d\omega \, \sum_P (-1)^P  \non
& \times \,
   \Bigg\{\,
     \langle \psi_{Pi_1} \psi_{Pi_2} | R(\omega-\veps_{Pi_1}) | \psi_{k_1} \psi_{k_2} \rangle
        \lb \frac{1}{\omega -\veps_{k_1} +i0} + \frac{1}{E_i^{(0)}-\omega -\veps_{k_2} +i0} \rb   \non
& \quad\, +
     \langle \psi_{Pi_1} \psi_{Pi_2} | R(\veps_{k_1}-\omega) | \psi_{k_1} \psi_{k_2} \rangle      
        \lb \frac{1}{\omega -\veps_{Pi_1} +i0} + \frac{1}{E_k^{(0)}-\omega -\veps_{Pi_2} +i0} \rb
   \,\Bigg\} \, . \non
\end{align}
The integration over $\omega$ can be performed using the standard identity ($\omega_1<0<\omega_2$):
\begin{align}
\label{eq:sokhot}
\int\limits_{\omega_1}^{\omega_2} \! d\omega \, \frac{f(\omega)}{\omega \pm i0} =
\mp i\pi f(0) + {\rm{P.V.}} \int\limits_{\omega_1}^{\omega_2} \! d\omega \, \frac{f(\omega)}{\omega} \, , 
\end{align}
where ${\rm{P.V}}$ means the principal-value integral. Indeed, applying the formula~(\ref{eq:sokhot}) to all four terms in Eq.~(\ref{eq:H_2el}) and taking into account that all the principal-value integrals vanish due to the fact that $R(\omega)$ is the even function of $\omega$, one finally obtains the SMS contribution 
\begin{align}
\label{eq:H_2el_total}
H_{{\rm SMS}, ik}
= \frac{1}{2}
\sum_P (-1)^P \Big[ 
\langle \psi_{Pi_1} \psi_{Pi_2} | R(\Delta_{1}) | \psi_{k_1} \psi_{k_2} \rangle
+
\langle \psi_{Pi_1} \psi_{Pi_2} | R(\Delta_{2}) | \psi_{k_1} \psi_{k_2} \rangle
\Big] \, ,
\end{align}
where $\Delta_{1} = \veps_{Pi_1} - \veps_{k_1}$ and $\Delta_{2} = \veps_{Pi_2} - \veps_{k_2}$.

\section{\label{sec:b} Computational formulas for the one-electron contributions}

\begin{figure}[h]
\begin{center}
{}
\hfill
\includegraphics[width=0.46\textwidth]{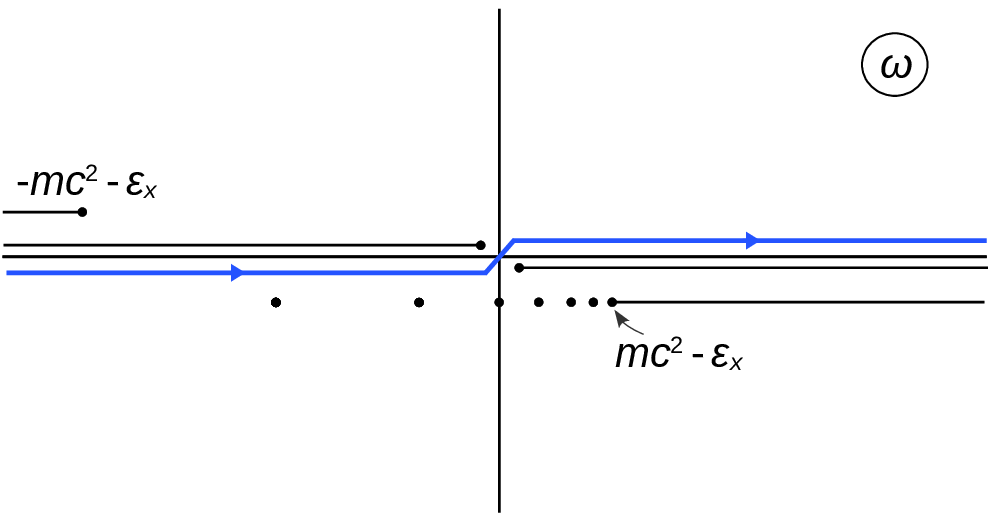}
\hfill
\includegraphics[width=0.46\textwidth]{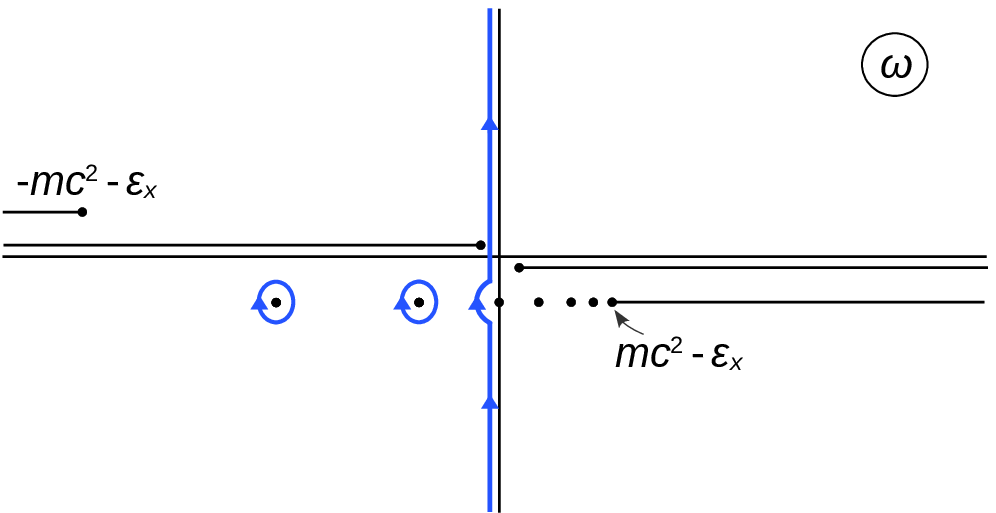}
{}
\caption{\label{fig:contour_pole}
The poles and the branch cuts of the integrand in the operator $\rP(\veps_x)$ for the one- and two-transverse-photon contributions. The integration contour: the original one oriented along the real axis (left panel); the rotated to the imaginary one (right panel).}
\end{center}
\end{figure}
\begin{figure}
\begin{center}
\includegraphics[width=0.46\textwidth]{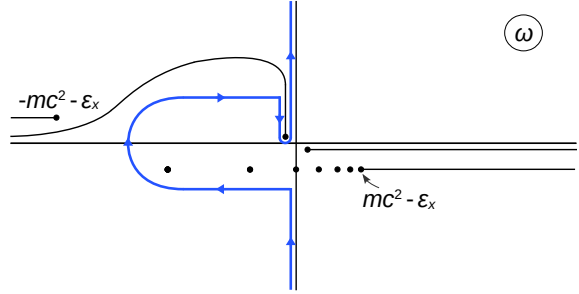}
\caption{\label{fig:contour_cmplx}
The poles and the branch cuts of the integrand in the operator $\rP(\veps_x)$ for the one- and two-transverse-photon contributions. The integration contour in the complex plane is chosen to avoid all the singularities of the integrand.
}
\end{center}
\end{figure}

In view of the definition~(\ref{eq:P_operator}), the expression~(\ref{eq:H_1el_total}) obtained in Appendix~\ref{sec:a} is given in a form that allows one to use the finite-basis-set methods~\cite{Johnson:1988:307, Salomonson:1989:5548, splines:DKB} for its calculations. Therefore, the main difficulty is related with the evaluation of the integral over $\omega$. For large real values of $\omega$, the photon propagator~(\ref{eq:D_lk}) is a strongly oscillating function. It is convenient to perform the Wick's rotation to overcome this obstacle. In the present Appendix, we discuss the related transformations for the contribution~$H_{{\rm NMS}, ik}$.

Employing Eq.~(\ref{eq:R_sum}), one can represent Eq.~(\ref{eq:H_1el_total}) as the sum of the Coulomb, one-transverse-photon, and two-transverse-photon contributions
\begin{align}
H_{{\rm NMS}, ik}
= H_{{\rm NMS}, ik}^{\rm c} + H_{{\rm NMS}, ik}^{\rm tr1} + H_{{\rm NMS}, ik}^{\rm tr2}\, .
\end{align}
In the Coulomb contribution, the integration over $\omega$ can be performed analytically by means of the identity~(\ref{eq:sokhot})
\begin{align}
\label{eq:omint_c}
H_{{\rm NMS}, ik}^{\rm c} = \frac{1}{2} \left[ \, \sum_n^{\veps_{n} > 0} \bra{\psi_{i} \psi_{n}} {R}_{\rm c} \ket{\psi_{n} \psi_{k}} - \sum_n^{\veps_{n} < 0} \bra{\psi_{i} \psi_{n}} {R}_{\rm c} \ket{\psi_{n} \psi_{k}} \right] \, .
\end{align}

The one- and two-transverse-photon contributions are handled identically. In what follows in this Appendix, we will refer to them together as the ``transverse-photon'' (tr) ones. In the left panel of Fig.~\ref{fig:contour_pole}, the original integration contour oriented along the real axis is shown. The poles and the branch cuts of the integrand are presented as well. Upon the Wick's rotation to the imaginary axis, the bound-state poles of the electron Green's function are picked up as the residues, see the right panel in Fig.~\ref{fig:contour_pole}. The final formulas for the transverse-photon contribution can be written as a sum of three terms
\begin{align}
H_{{\rm NMS}, ik}^{\rm tr} = H_{{\rm NMS}, ik}^{{\rm tr}(a)} + H_{{\rm NMS}, ik}^{{\rm tr}(b)} + H_{{\rm NMS}, ik}^{{\rm tr}(c)} \, .
\end{align}
The first term arises from the residues shown by the circles in Fig.~\ref{fig:contour_pole} and reads as
\begin{align}
\label{eq:omint_tr_a}
H_{{\rm NMS}, ik}^{{\rm tr}(a)} = \frac{1}{2} \sum_{x=i,k} \! \sum_n^{-1<\veps_{n} < \veps_{x}} \!
\bra{\psi_{i} \psi_{n}} {R}_{\rm tr}(\veps_{n} - \veps_{x}) \ket{\psi_{n} \psi_{k}} \, ,
\end{align}
where for brevity we have introduced the summation over $x$ in order to take into account the symmetric form of the expression with respect to the argument of $\rP(E)$. The second term corresponds to the case of degeneracy between the states of the opposite parity. This term originates from the poles located at $\omega=0$ and has the form  
\begin{align}
\label{eq:omint_tr_b}
H_{{\rm NMS}, ik}^{{\rm tr}(b)} = \frac{1}{4} \sum_{x=i,k}  \sum_n^{\veps_{n} = \veps_{x}} \bra{\psi_{i} \psi_{n}} {R}_{\rm tr}(0) \ket{\psi_{n} \psi_{k}}\, .
\end{align}
Finally, the third term corresponds to the integration over the imaginary axis
\begin{align}
\label{eq:omint_tr_c}
H_{{\rm NMS}, ik}^{{\rm tr}(c)} = \frac{1}{2} \sum_{x=i,k}  \sum_n^{\veps_{n} \neq \veps_{x}} \frac{1}{\pi} \int_{0}^{\infty} dy \,\, \frac{\veps_{n} - \veps_{x}}{y^{2} + (\veps_{n}- \veps_{x})^{2}}  \bra{\psi_{i} \psi_{n}} {R}_{\rm tr}(iy) \ket{\psi_{n} \psi_{k}} \, .
\end{align}

The expressions similar to (\ref{eq:omint_c}), (\ref{eq:omint_tr_a}), (\ref{eq:omint_tr_b}), and (\ref{eq:omint_tr_c}) were derived, e.g., in Ref.~\cite{Artemyev:1995:1884}. However, only the case of the diagonal matrix elements was considered. Moreover, in Ref.~\cite{Artemyev:1995:1884} the formulas for the higher-order QED corrections beyond the Breit approximation were given, while the present ones include the lowest-relativistic contributions as well.

As an additional crosscheck of the $\omega$-integration routine, we perform it also by employing the contour schematically shown in Fig.~\ref{fig:contour_cmplx}. This contour is chosen to bypass all the singularities in the complex plane. The matrix elements~$H_{{\rm NMS}, ik}$ 
have been evaluated for both variants of the integration-contour rotation. The results are found to be in excellent agreement with each other.

\section{\label{sec:c} Local effective potentials}

In Sec.~\ref{sec:3}, in order to test the performance of the model-QED operator for the nuclear recoil effect, we need substitute the core-Hartree (CH) and $x_\alpha$ potentials (see, e.g., Ref.~\cite{Sapirstein:2002:042501}) instead of $V$ in Eq.~(\ref{eq:Dirac_eq}). These potentials can be expressed via the charge densities of the valence electron,
\begin{align}
\label{eq:dense_v}
\rho_v(r) = g_v^2(r) + f_v^2(r) \, ,
\end{align}
and the core, 
\begin{align}
\label{eq:dense_c}
\rho_{\rm core}(r) = \sum_c (2j_c+1) \Big[ g_c^2(r) + f_c^2(r) \Big] \, .
\end{align}
Here the large, $g$, and small, $f$, components of the Dirac wave function in Eqs.~(\ref{eq:dense_v}) and (\ref{eq:dense_c}) are determined self-consistently in the local potential
\begin{align}
\label{eq:V}
V(r) = -\frac{\alpha Z_{\rm eff}(r)}{r} \, ,
\end{align}
where $Z_{\rm eff}(r)$ is the effective charge. The CH potential is given by
\begin{align}
\label{eq:CH}
Z_{\rm eff}^{\rm CH}(r) = 
Z_{\rm nucl}(r) - r \int\limits_0^\infty \! dr' \, \frac{\rho_{\rm core}(r')}{{\rm max}\{r,r'\}} \, ,
\end{align}
where $Z_{\rm nucl}(r)$ accounts for the finite size of the nucleus. Defining the total charge density, $\rho_{\rm tot}=\rho_{\rm core}+\rho_v$, the effective charge for the $x_\alpha$ potentials can be written in the form
\begin{align}
\label{eq:x_alpha}
Z_{\rm eff}^{x_\alpha}(r) = 
Z_{\rm nucl}(r) - r \int\limits_0^\infty \! dr' \, \frac{\rho_{\rm tot}(r')}{{\rm max}\{r,r'\}}
+ x_\alpha \left[ \frac{81}{32\pi^2} r \rho_{\rm tot}(r) \right]^{1/3} \, .
\end{align}
We use the values of $x_\alpha$ equal to 0, 1/3, 2/3, and 1. The choice $x_\alpha=0$ leads to the Dirac-Hartree potential, $x_\alpha=2/3$ corresponds to the Kohn-Sham potential~\cite{pot:KS}, while $x_\alpha=1$ is the Dirac-Slater potential~\cite{Slater:1951:385}. In order to restore the proper asymptotic behavior of the $x_\alpha$ potentials, we introduce the Latter correction~\cite{Latter:1955:510}.



\end{document}

%% file: ns_nondiag.tex
{
\begin{table}
\centering
\renewcommand{\arraystretch}{0.608}
\renewcommand{\baselinestretch}{0.69}
\caption{\label{tab:ns_nondiag} Matrix elements of the operator~$h^{\rm NMS}_{\rm h.o.}$ for the $ns$ states calculated with the Dirac-Coulomb wave functions for the extended nuclei. Labels~$(n_i,n_k)$ stand for the function $F_{n_in_k}$ defined by Eq.~(\ref{eq:F_ik}). Rows labeled ``\rm{c}'', ``\rm{tr1}'', ``\rm{tr2}'', and ``\rm{tot}'' contain the Coulomb, one-transverse-photon, two-transverse-photon and total nuclear recoil contributions, respectively.}
\resizebox{\textwidth}{!}{ 
\begin{tabular}{@{\extracolsep{\fill}}
                 c
                 c|*
                 {15}{S[table-format=-1.5]}
                 }

\hline
\hline
 \multicolumn{1}{c}{$Z$} & \multicolumn{1}{c|}{Term} & \multicolumn{1}{c}{$(1,1)$} & \multicolumn{1}{c}{$(1,2)$} & \multicolumn{1}{c}{$(1,3)$} & \multicolumn{1}{c}{$(1,4)$} & \multicolumn{1}{c}{$(1,5)$} & \multicolumn{1}{c}{$(2,2)$}& \multicolumn{1}{c}{$(2,3)$} & \multicolumn{1}{c}{$(2,4)$}& \multicolumn{1}{c}{$(2,5)$} & \multicolumn{1}{c}{$(3,3)$} & \multicolumn{1}{c}{$(3,4)$} & \multicolumn{1}{c}{$(3,5)$} & \multicolumn{1}{c}{$(4,4)$} & \multicolumn{1}{c}{$(4,5)$} & \multicolumn{1}{c}{$(5,5)$} \\
\hline

     5  &  \rm{c}  &    -0.38522  &    -0.38532  &    -0.38531  &    -0.38530  &    -0.38528  &    -0.38542  &    -0.38540  &    -0.38539  &    -0.38538  &    -0.38539  &    -0.38538  &    -0.38537  &    -0.38537  &    -0.38536  &    -0.38535    \\ 
     &  \rm{tr1}   &     2.73291  &     2.80247  &     2.80794  &     2.80923  &     2.80968  &     2.92571  &     2.94192  &     2.94448  &     2.94525  &     2.97187  &     2.97825  &     2.97968  &     2.98997  &     2.99314  &     2.99890    \\ 
     &  \rm{tr2}   &    -0.97002  &    -0.93991  &    -0.93526  &    -0.93365  &    -0.93291  &    -0.93021  &    -0.92415  &    -0.92237  &    -0.92156  &    -0.92221  &    -0.92003  &    -0.91917  &    -0.91932  &    -0.91830  &    -0.91796    \\ 
     &  \rm{tot}   &     1.37767  &     1.47725  &     1.48737  &     1.49028  &     1.49149  &     1.61008  &     1.63236  &     1.63672  &     1.63831  &     1.66426  &     1.67284  &     1.67514  &     1.68528  &     1.68949  &     1.69560    \\ 
 \hline

   10  &  \rm{c}  &    -0.35934  &    -0.35973  &    -0.35970  &    -0.35965  &    -0.35961  &    -0.36014  &    -0.36010  &    -0.36006  &    -0.36002  &    -0.36007  &    -0.36002  &    -0.35999  &    -0.35998  &    -0.35994  &    -0.35990    \\
\hline
\hline
\end{tabular}
}
\end{table}
}

%% file: np1_nondiag.tex
{
\begin{table} 
\centering
\renewcommand{\arraystretch}{0.595}
\renewcommand{\baselinestretch}{0.70}
\caption{\label{tab:np1_nondiag} Matrix elements of the operator~$h^{\rm NMS}_{\rm h.o.}$ for the $np_{1/2}$ states calculated with the Dirac-Coulomb wave functions for the extended nuclei. The notations are the same as in Table~\ref{tab:ns_nondiag}.}
\resizebox{0.725\textwidth}{!}{
\begin{tabular}{@{\extracolsep{\fill}}
                 c
                 c|*
                 {10}{S[table-format=-1.5]}
                 }
\hline
\hline
 \multicolumn{1}{c}{$Z$} & \multicolumn{1}{c|}{Term} & \multicolumn{1}{c}{$(2,2)$}& \multicolumn{1}{c}{$(2,3)$} & \multicolumn{1}{c}{$(2,4)$}& \multicolumn{1}{c}{$(2,5)$} & \multicolumn{1}{c}{$(3,3)$} & \multicolumn{1}{c}{$(3,4)$} & \multicolumn{1}{c}{$(3,5)$} & \multicolumn{1}{c}{$(4,4)$} & \multicolumn{1}{c}{$(4,5)$} & \multicolumn{1}{c}{$(5,5)$} \\
\hline

     5  &  \rm{c}  &    -0.00021  &    -0.00022  &    -0.00023  &    -0.00023  &    -0.00024  &    -0.00025  &    -0.00025  &    -0.00026  &    -0.00026  &    -0.00026    \\ 
     &  \rm{tr1}   &    -0.04748  &    -0.04734  &    -0.04921  &    -0.05021  &    -0.04134  &    -0.04170  &    -0.04296  &    -0.03843  &    -0.03863  &    -0.03679    \\ 
     &  \rm{tr2}   &    -0.03802  &    -0.03395  &    -0.03302  &    -0.03264  &    -0.03570  &    -0.03442  &    -0.03400  &    -0.03489  &    -0.03432  &    -0.03451    \\ 
     &  \rm{tot}   &    -0.08570  &    -0.08151  &    -0.08246  &    -0.08308  &    -0.07729  &    -0.07637  &    -0.07720  &    -0.07357  &    -0.07320  &    -0.07156    \\ 
 \hline

   10  &  \rm{c}  &    -0.00076  &    -0.00083  &    -0.00085  &    -0.00086  &    -0.00090  &    -0.00092  &    -0.00093  &    -0.00095  &    -0.00096  &    -0.00097    \\ 
    &  \rm{tr1}   &    -0.04858  &    -0.04867  &    -0.05059  &    -0.05162  &    -0.04294  &    -0.04337  &    -0.04465  &    -0.04019  &    -0.04042  &    -0.03863    \\ 
    &  \rm{tr2}   &    -0.02313  &    -0.01817  &    -0.01702  &    -0.01654  &    -0.01855  &    -0.01693  &    -0.01639  &    -0.01695  &    -0.01622  &    -0.01620    \\
\hline
\hline
\end{tabular}
}
\end{table}
}

%% file: np3_nondiag.tex
{\begin{table}
\centering
\renewcommand{\arraystretch}{0.595}
\renewcommand{\baselinestretch}{0.70}
\caption{\label{tab:np3_nondiag} Matrix elements of the operator~$h^{\rm NMS}_{\rm h.o.}$ for the $np_{3/2}$ states calculated with the Dirac-Coulomb wave functions for the extended nuclei. The notations are the same as in Table~\ref{tab:ns_nondiag}.}
\resizebox{0.725\textwidth}{!}{ 
\begin{tabular}{
                 c
                 c|*
                 {10}{S[table-format=-1.5]}
                 }
 
\hline
\hline
 \multicolumn{1}{c}{$Z$} & \multicolumn{1}{c|}{Term} & \multicolumn{1}{c}{$(2,2)$}& \multicolumn{1}{c}{$(2,3)$} & \multicolumn{1}{c}{$(2,4)$}& \multicolumn{1}{c}{$(2,5)$} & \multicolumn{1}{c}{$(3,3)$} & \multicolumn{1}{c}{$(3,4)$} & \multicolumn{1}{c}{$(3,5)$} & \multicolumn{1}{c}{$(4,4)$} & \multicolumn{1}{c}{$(4,5)$} & \multicolumn{1}{c}{$(5,5)$} \\
\hline

     5  &  \rm{c}  &    -0.00004  &    -0.00004  &    -0.00005  &    -0.00005  &    -0.00005  &    -0.00005  &    -0.00005  &    -0.00005  &    -0.00005  &    -0.00005    \\ 
     &  \rm{tr1}   &    -0.03916  &    -0.03865  &    -0.04038  &    -0.04132  &    -0.03221  &    -0.03243  &    -0.03362  &    -0.02901  &    -0.02915  &    -0.02724    \\ 
     &  \rm{tr2}   &    -0.04751  &    -0.04389  &    -0.04311  &    -0.04279  &    -0.04628  &    -0.04517  &    -0.04483  &    -0.04585  &    -0.04536  &    -0.04565    \\ 
     &  \rm{tot}   &    -0.08671  &    -0.08259  &    -0.08354  &    -0.08416  &    -0.07854  &    -0.07765  &    -0.07850  &    -0.07491  &    -0.07456  &    -0.07294    \\ 
 \hline

   10  &  \rm{c}  &    -0.00013  &    -0.00015  &    -0.00015  &    -0.00015  &    -0.00016  &    -0.00016  &    -0.00016  &    -0.00017  &    -0.00017  &    -0.00017    \\ 
    &  \rm{tr1}   &    -0.03300  &    -0.03250  &    -0.03421  &    -0.03513  &    -0.02601  &    -0.02622  &    -0.02740  &    -0.02279  &    -0.02292  &    -0.02101    \\ 
    &  \rm{tr2}   &    -0.04260  &    -0.03859  &    -0.03774  &    -0.03740  &    -0.04029  &    -0.03904  &    -0.03865  &    -0.03947  &    -0.03892  &    -0.03909    \\
\hline
\hline
\end{tabular}
}
\end{table}
}

%% file: nd3_nondiag.tex
{
\begin{table} 
\centering
\renewcommand{\arraystretch}{0.595}
\renewcommand{\baselinestretch}{0.70}
\caption{\label{tab:nd3_nondiag} Matrix elements of the operator~$h^{\rm NMS}_{\rm h.o.}$ for the $nd_{3/2}$ states calculated with the Dirac-Coulomb wave functions for the extended nuclei. The notations are the same as in Table~\ref{tab:ns_nondiag}.}
\resizebox{0.464\textwidth}{!}{
\begin{tabular}{@{\extracolsep{\fill}}
                 c
                 c|*
                 {6}{S[table-format=-1.5]}
                 }

\hline
\hline
 \multicolumn{1}{c}{$Z$} & \multicolumn{1}{c|}{Term} & \multicolumn{1}{c}{$(3,3)$} & \multicolumn{1}{c}{$(3,4)$} & \multicolumn{1}{c}{$(3,5)$} & \multicolumn{1}{c}{$(4,4)$} & \multicolumn{1}{c}{$(4,5)$} & \multicolumn{1}{c}{$(5,5)$} \\
\hline

     5  &  \rm{c}  &    -0.00000  &    -0.00000  &    -0.00000  &    -0.00000  &    -0.00000  &    -0.00000    \\ 
     &  \rm{tr1}   &    -0.00932  &    -0.00918  &    -0.00969  &    -0.00814  &    -0.00840  &    -0.00746    \\ 
     &  \rm{tr2}   &    -0.00938  &    -0.00796  &    -0.00756  &    -0.00908  &    -0.00853  &    -0.00894    \\ 
     &  \rm{tot}   &    -0.01870  &    -0.01713  &    -0.01726  &    -0.01722  &    -0.01693  &    -0.01639    \\ 
 \hline

   10  &  \rm{c}  &    -0.00000  &    -0.00000  &    -0.00000  &    -0.00000  &    -0.00000  &    -0.00000    \\ 
    &  \rm{tr1}   &    -0.00896  &    -0.00889  &    -0.00944  &    -0.00789  &    -0.00817  &    -0.00725    \\ 
    &  \rm{tr2}   &    -0.00814  &    -0.00664  &    -0.00624  &    -0.00753  &    -0.00694  &    -0.00724    \\
\hline
\hline
\end{tabular}
}
\end{table}
}

%% file: nd5_nondiag.tex
{
\begin{table} 
\centering
\renewcommand{\arraystretch}{0.595}
\renewcommand{\baselinestretch}{0.70}
\caption{\label{tab:nd5_nondiag} Matrix elements of the operator~$h^{\rm NMS}_{\rm h.o.}$ for the $nd_{5/2}$ states calculated with the Dirac-Coulomb wave functions for the extended nuclei. The notations are the same as in Table~\ref{tab:ns_nondiag}.}
\resizebox{0.464\textwidth}{!}{
\begin{tabular}{@{\extracolsep{\fill}}
                 c
                 c|*
                 {6}{S[table-format=-1.5]}
                 }
\hline
\hline
 \multicolumn{1}{c}{$Z$} & \multicolumn{1}{c|}{Term} & \multicolumn{1}{c}{$(3,3)$} & \multicolumn{1}{c}{$(3,4)$} & \multicolumn{1}{c}{$(3,5)$} & \multicolumn{1}{c}{$(4,4)$} & \multicolumn{1}{c}{$(4,5)$} & \multicolumn{1}{c}{$(5,5)$} \\
\hline

     5  &  \rm{c}  &    -0.00000  &    -0.00000  &    -0.00000  &    -0.00000  &    -0.00000  &    -0.00000    \\ 
     &  \rm{tr1}   &    -0.00863  &    -0.00846  &    -0.00896  &    -0.00737  &    -0.00761  &    -0.00665    \\ 
     &  \rm{tr2}   &    -0.01006  &    -0.00865  &    -0.00826  &    -0.00984  &    -0.00931  &    -0.00974    \\ 
     &  \rm{tot}   &    -0.01869  &    -0.01711  &    -0.01722  &    -0.01721  &    -0.01692  &    -0.01639    \\ 
 \hline

   10  &  \rm{c}  &    -0.00000  &    -0.00000  &    -0.00000  &    -0.00000  &    -0.00000  &    -0.00000    \\ 
    &  \rm{tr1}   &    -0.00756  &    -0.00744  &    -0.00794  &    -0.00632  &    -0.00656  &    -0.00560    \\ 
    &  \rm{tr2}   &    -0.00951  &    -0.00804  &    -0.00766  &    -0.00907  &    -0.00851  &    -0.00887    \\
\hline
\hline
\end{tabular}
}
\end{table}
}

%% file: recmod_test.tex
{
\begin{table}
\centering
\renewcommand{\arraystretch}{0.9}
\renewcommand{\baselinestretch}{0.90}
\caption{\label{tab:recmod_predict} The higher-order nuclear recoil correction for the $4s$, $5s$, $5p$, and $5d$ states in terms of the function $F$ defined by Eq.~(\ref{eq:F_ik}). The column labeled $\langle\psi_v| \tilde{h}^{\rm NMS}_{\rm h.o.} |\psi_v \rangle$ denotes the results obtained by means of the model QED operator, $\langle\psi_v| V_{\rm s.l.} |\psi_v \rangle$ is the contribution of the semilocal part of the model operator, and ``Exact'' stands for the \textit{ab initio} values. The $ns$ states with $n \leqslant 3$ and the $np$ and $nd$ states with $n \leqslant 4$ are omitted since the operator $\tilde{h}^{\rm NMS}_{\rm h.o.}$ exactly reproduces the corresponding correction for them by construction. }
\resizebox{\textwidth}{!}{ 
\begin{tabular}{l@{\quad}
                l@{\quad}
                S[table-format=-2.5]
                S[table-format=-2.5]
                S[table-format=-2.5]
                r@{\quad}
                l@{\quad}
                S[table-format=-2.5]
                S[table-format=-2.5]
                S[table-format=-2.5]
                 }

\hline
\hline
  $Z$ & State &  
  \multicolumn{1}{c}{$\langle\psi_v| V_{\rm s.l.} |\psi_v \rangle$} & 
  \multicolumn{1}{c}{$\langle\psi_v| \tilde{h}^{\rm NMS}_{\rm h.o.} |\psi_v \rangle$}  & 
  \multicolumn{1}{c}{Exact} &
  $~~~~Z$ & State &  
  \multicolumn{1}{c}{$\langle\psi_v| V_{\rm s.l.} |\psi_v \rangle$} & 
  \multicolumn{1}{c}{$\langle\psi_v| \tilde{h}^{\rm NMS}_{\rm h.o.} |\psi_v \rangle$}  & 
  \multicolumn{1}{c}{Exact} \\
\hline

 10 & $4s$       &   1.2024  &   1.5325  &   1.5295  &   60 & $4s$       &   0.8494  &  1.4804  &  1.4767  \\
    & $5s$       &   1.2017  &   1.5445  &   1.5397  &      & $5s$       &   0.8394  &  1.4833  &  1.4772  \\
    & $5p_{1/2}$ &  -0.0922  &  -0.0545  &  -0.0558  &      & $5p_{1/2}$ &   0.1282  &  0.1782  &  0.1765  \\
    & $5p_{3/2}$ &  -0.0965  &  -0.0590  &  -0.0603  &      & $5p_{3/2}$ &   0.0221  &  0.0540  &  0.0517  \\
    & $5d_{3/2}$ &  -0.0279  &  -0.0157  &  -0.0145  &      & $5d_{3/2}$ &  -0.0010  &  0.0058  &  0.0053  \\
    & $5d_{5/2}$ &  -0.0278  &  -0.0156  &  -0.0145  &      & $5d_{5/2}$ &  -0.0019  &  0.0044  &  0.0043  \\
\hline                          
                                
 20 & $4s$       &   1.0162  &   1.3896  &   1.3865  &   70 & $4s$       &   0.9227  &  1.6503  &  1.6463  \\
    & $5s$       &   1.0141  &   1.4010  &   1.3960  &      & $5s$       &   0.9088  &  1.6475  &  1.6409  \\
    & $5p_{1/2}$ &  -0.0555  &  -0.0204  &  -0.0218  &      & $5p_{1/2}$ &   0.1963  &  0.2652  &  0.2634  \\
    & $5p_{3/2}$ &  -0.0692  &  -0.0349  &  -0.0363  &      & $5p_{3/2}$ &   0.0406  &  0.0762  &  0.0736  \\
    & $5d_{3/2}$ &  -0.0225  &  -0.0116  &  -0.0107  &      & $5d_{3/2}$ &   0.0047  &  0.0110  &  0.0101  \\
    & $5d_{5/2}$ &  -0.0224  &  -0.0116  &  -0.0107  &      & $5d_{5/2}$ &   0.0028  &  0.0084  &  0.0080  \\
\hline                                                  
                                                        
 30 & $4s$       &   0.9090  &   1.3357  &   1.3325  &   80 & $4s$       &   1.0683  &  1.9249  &  1.9205  \\
    & $5s$       &   0.9053  &   1.3460  &   1.3408  &      & $5s$       &   1.0478  &  1.9125  &  1.9055  \\
    & $5p_{1/2}$ &  -0.0162  &   0.0177  &   0.0162  &      & $5p_{1/2}$ &   0.2880  &  0.3904  &  0.3885  \\
    & $5p_{3/2}$ &  -0.0437  &  -0.0119  &  -0.0136  &      & $5p_{3/2}$ &   0.0578  &  0.0990  &  0.0961  \\
    & $5d_{3/2}$ &  -0.0172  &  -0.0075  &  -0.0069  &      & $5d_{3/2}$ &   0.0108  &  0.0169  &  0.0156  \\
    & $5d_{5/2}$ &  -0.0171  &  -0.0076  &  -0.0069  &      & $5d_{5/2}$ &   0.0073  &  0.0124  &  0.0117  \\
\hline                                                  
                                                        
 40 & $4s$       &   0.8484  &   1.3348  &   1.3314  &   90 & $4s$       &   1.3278  &  2.3698  &  2.3650  \\
    & $5s$       &   0.8430  &   1.3436  &   1.3381  &      & $5s$       &   1.2952  &  2.3406  &  2.3329  \\
    & $5p_{1/2}$ &   0.0262  &   0.0612  &   0.0596  &      & $5p_{1/2}$ &   0.4244  &  0.5875  &  0.5855  \\
    & $5p_{3/2}$ &  -0.0200  &   0.0103  &   0.0084  &      & $5p_{3/2}$ &   0.0739  &  0.1228  &  0.1195  \\
    & $5d_{3/2}$ &  -0.0119  &  -0.0033  &  -0.0031  &      & $5d_{3/2}$ &   0.0173  &  0.0236  &  0.0220  \\
    & $5d_{5/2}$ &  -0.0119  &  -0.0036  &  -0.0032  &      & $5d_{5/2}$ &   0.0117  &  0.0165  &  0.0155  \\
\hline                                                  
                                                        
 50 & $4s$       &   0.8280  &   1.3813  &   1.3777  &  100 & $4s$       &   1.7885  &  3.1190  &  3.1142  \\ 
    & $5s$       &   0.8206  &   1.3877  &   1.3820  &      & $5s$       &   1.7322  &  3.0568  &  3.0496  \\ 
    & $5p_{1/2}$ &   0.0733  &   0.1131  &   0.1115  &      & $5p_{1/2}$ &   0.6515  &  0.9319  &  0.9297  \\ 
    & $5p_{3/2}$ &   0.0019  &   0.0322  &   0.0301  &      & $5p_{3/2}$ &   0.0889  &  0.1478  &  0.1444  \\ 
    & $5d_{3/2}$ &  -0.0065  &   0.0011  &   0.0010  &      & $5d_{3/2}$ &   0.0246  &  0.0317  &  0.0297  \\ 
    & $5d_{5/2}$ &  -0.0068  &   0.0004  &   0.0006  &      & $5d_{5/2}$ &   0.0158  &  0.0206  &  0.0193  \\  

\hline
\hline
\end{tabular}
}
\end{table}
}

%% file: recmod_alkali.tex
{
\begin{table}
\centering
\renewcommand{\arraystretch}{0.9}
\renewcommand{\baselinestretch}{0.90}
\caption{\label{tab:recmod_alkali} The one-electron nuclear recoil correction beyond the Breit approximation for the valence $ns$ electron in neutral alkali metals in terms of the function $F$ defined by Eq.~(\ref{eq:F_ik}). The labels CH and $x_\alpha=0$, 1/3, 2/3, and 1 correspond to the different effective potentials. See the text for details.}
%
\begin{tabular}{l@{\quad}
                c@{\quad}
                @{\quad}S[table-format=2.6,group-separator=]
                @{\quad}S[table-format=2.6,group-separator=]
                @{\quad}S[table-format=2.6,group-separator=]
                @{\quad}S[table-format=2.6,group-separator=]
                @{\quad}S[table-format=2.6,group-separator=]
                 }

\hline
\hline
  Atom & Approach &  
  \multicolumn{1}{c}{CH} & 
  \multicolumn{1}{c}{$x_\alpha=0$}  & 
  \multicolumn{1}{c}{$x_\alpha=1/3$}  & 
  \multicolumn{1}{c}{$x_\alpha=2/3$}  & 
  \multicolumn{1}{c}{$x_\alpha=1$}  \\
\hline

 {\rm Na $3s$} &  $\langle\psi_v| V_{\rm s.l.} |\psi_v \rangle$                    &  0.0502  &   0.0446  &   0.0437  &   0.0473  &   0.0575  \\ 
               &  $\langle\psi_v| \tilde{h}^{\rm NMS}_{\rm h.o.} |\psi_v \rangle$  &  0.0581  &   0.0516  &   0.0508  &   0.0553  &   0.0676  \\
               &  Exact                                                            &  0.0561  &   0.0499  &   0.0496  &   0.0544  &   0.0671  \\
               
\hline  

 {\rm K $4s$}  &  $\langle\psi_v| V_{\rm s.l.} |\psi_v \rangle$                    &  0.0255  &   0.0214  &   0.0213  &   0.0243  &   0.0319  \\ 
               &  $\langle\psi_v| \tilde{h}^{\rm NMS}_{\rm h.o.} |\psi_v \rangle$  &  0.0313  &   0.0263  &   0.0264  &   0.0302  &   0.0400  \\
               &  Exact                                                            &  0.0311  &   0.0261  &   0.0263  &   0.0302  &   0.0401  \\
               
\hline    

 {\rm Rb $5s$} &  $\langle\psi_v| V_{\rm s.l.} |\psi_v \rangle$                    &  0.0100  &   0.0080  &   0.0082  &   0.0098  &   0.0136  \\ 
               &  $\langle\psi_v| \tilde{h}^{\rm NMS}_{\rm h.o.} |\psi_v \rangle$  &  0.0141  &   0.0112  &   0.0116  &   0.0140  &   0.0195  \\
               &  Exact                                                            &  0.0142  &   0.0113  &   0.0117  &   0.0141  &   0.0197  \\
               
\hline    

 {\rm Cs $6s$} &  $\langle\psi_v| V_{\rm s.l.} |\psi_v \rangle$                    &  0.00645  &  0.00504  &  0.00525  &  0.00642  &  0.00927  \\
               &  $\langle\psi_v| \tilde{h}^{\rm NMS}_{\rm h.o.} |\psi_v \rangle$  &  0.01019  &  0.00796  &  0.00835  &  0.01026  &  0.01490  \\
               &  Exact                                                            &  0.01028  &  0.00803  &  0.00841  &  0.01034  &  0.01500  \\
               
\hline    

 {\rm Fr $7s$} &  $\langle\psi_v| V_{\rm s.l.} |\psi_v \rangle$                    &  0.00586  &  0.00416  &  0.00457  &  0.00595  &  0.00906  \\
               &  $\langle\psi_v| \tilde{h}^{\rm NMS}_{\rm h.o.} |\psi_v \rangle$  &  0.00999  &  0.00708  &  0.00782  &  0.01022  &  0.01563  \\
               &  Exact                                                            &  0.01004  &  0.00712  &  0.00786  &  0.01026  &  0.01568  \\

\hline
\hline
\end{tabular}
\end{table}
}